# Topological Hall Effect in Antiferromagnetic Co doped $Fe_3GaTe_2$

*Shyam Raj Karullithodi[1, 2, ‡], Yeonkyu Lee[3, 4, ‡], Vadym Kulichenko[1], W. Kice Brown[5], Sang-Eon Lee[1], Chanyoung Lee[3, 4], Jinyoung Yun[3, 4], Gregory T. McCandless[5], Julia Y. Chan[5], Jeehoon Kim[3, 4, *], and Luis Balicas[6, *]*

1. National High Magnetic Field Laboratory, Tallahassee, Florida 32310, United States
2. Department of Physics, Florida State University, Tallahassee, Florida 32306, United States
3. Department of Physics, Pohang University of Science and Technology, Pohang 37673, Republic of Korea
4. Center for Quantum Dynamics of Angular Momentum, Pohang University of Science and Technology, Pohang 37673, Republic of Korea
5. Department of Chemistry & Biochemistry, Baylor University, Waco, Texas 76706, United States
6. Department of Physics and Astronomy, Baylor University, Waco, Texas 76706, United States

*Correspondence to E-mail: jeehoon@postech.ac.kr; luis_balicas@baylor.edu

**ABSTRACT**: $Fe_3GaTe_2$ is van der Waals (vdW) ferromagnet with a Curie temperature $T_C$ ranging from 350 K to 380 K, followed upon cooling by a ferrimagnetic transition near room temperature. Substituting Fe with Co was previously reported to induce antiferromagnetism (AFM) at a Co fraction dependent Néel temperature $T_N$. In this work, we confirm the overall phase diagram of the $Fe_{3-x}Co_xGaTe_2$ series as a function of $x$ and temperature via magnetization and electrical transport measurements. For $x \geq 0.6$ the Hall effect is observed to mimic the magnetization as the AF ground state is suppressed by the external magnetic field via a metamagnetic transition, thus displaying an anomalous Hall response. At low temperatures, we also observe a pronounced topological Hall signal peaking at $\mu_0 H = 4$ T, or within the metamagnetic transition region of fields. This observation points to the presence of magnetic field-induced chiral spin textures, such as skyrmions upon approaching magnetization saturation. Magnetic force microscopy (MFM) reveals the emergence of nearly circular magnetic domains, with diameters on the order of 100–200 nm, within the antiferromagnetic phase. A detailed analysis of the MFM images indicates that the topological Hall effect is closely linked to the field-induced stabilization of magnetic domain structures, likely exhibiting chiral textures. This observation suggests the possible formation of skyrmions already in the AFM phase, i.e., AFM skyrmions, that evolve into ferromagnetic (FM) ones upon increasing the magnetic field. Consequently, Co-doped $Fe_3GaTe_2$ might provide a platform to investigate the transformation of skyrmions, initially coupled antiferromagnetically into ferromagnetic skyrmions, and to explore its impact on the topological and skyrmion Hall effects.

TOC Fig.

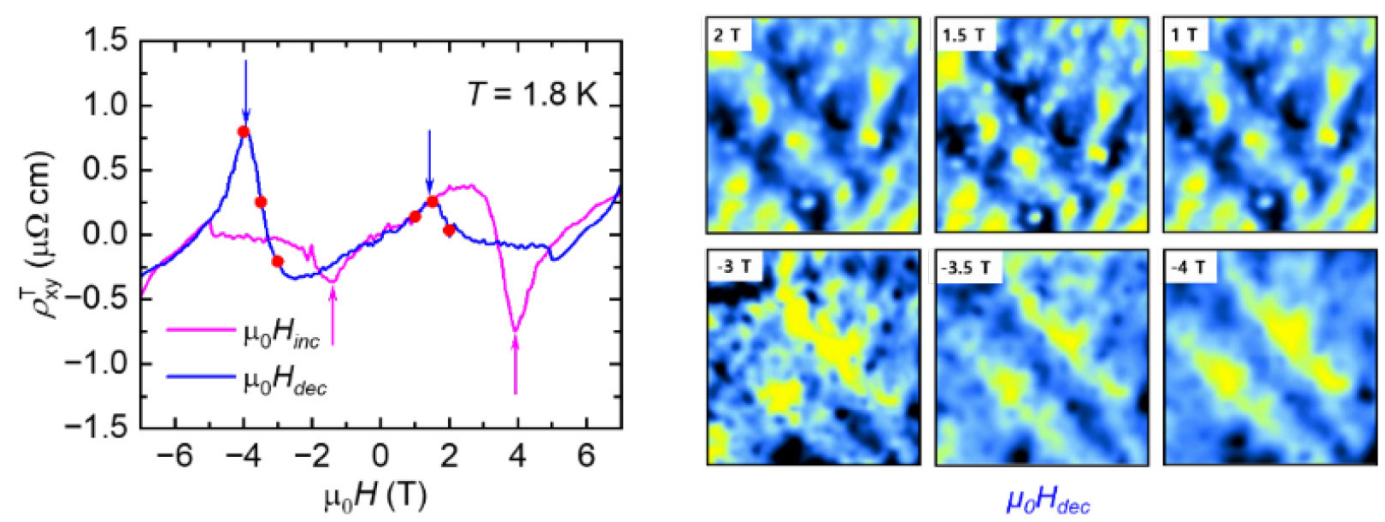

**TOC Fig.** Left panel: topological Hall effect $\rho_{xy}^{\mathrm{T}}$ as a function of magnetic field $\mu_0 H$ for field increasing (blue trace) and decreasing sweeps (magenta trace). Red markers indicate field and $\rho_{xy}^{\mathrm{T}}$values associated to the magnetic force microscopy images or domain structures observed in the right panel.

The search for ferromagnets with high Curie temperatures ($T_C$) and an inherent tendency to host chiral spin textures, such as skyrmions, is critical for advancing the field of skyrmionics.[1] The emergence of two-dimensional layered magnets, particularly those coupled by van der Waals (vdW) interactions, has opened new avenues for exploration. Several Fe based compounds in this class exhibit topological spin textures,[2-9] although such textures have also been theoretically proposed to form in twisted magnetic layers.[10, 11] Among vdW magnets, the layered $Fe_3GaTe_2$ is a unique ferromagnet since it displays high Curie temperatures $T_c$ ranging between $T_c \cong 350$ K and 380 K, higher than those of the other known vdW magnets while also displaying a large perpendicular magnetic anisotropy (PMA).[12] Given that itinerant ferromagnetism in $Fe_3GaTe_2$ persists down to the monolayer limit, its combination of large perpendicular magnetic anisotropy, high Curie temperature, and reported presence of topological spin textures[5-9] positions this material as a highly promising candidate for spintronic applications. Recently, we found evidence for a hysteretic phase transition from ferromagnetism to a ferrimagnetic ground state in $Fe_3GaTe_2$, that affects the density of the chiral spin textures observed in this compound as a function of the temperature.[13] This observation suggests that a Fe sub-lattice may couple AFM with the predominant ferromagnetically coupled Fe atoms.

Akin to ferromagnets, antiferromagnets can also host skyrmions,[14, 15] which are topologically protected particles but with greater potential for applications when they emerge in antiferromagnets relative to ferromagnets due to their zero net topological charge and hence zero skyrmion Hall effect. The dynamics of ferromagnetic skyrmion tracks are perturbed by the skyrmion Hall effect,[16] which leads to a motion transverse to the driving force or electrical current. This effect results from a Magnus force $Gz \times v$ acting on the moving skyrmion, a direct consequence of its topology (where $G$ is the amplitude of the skyrmion gyrovector which is proportional to the skyrmion topological charge $Q$ and the layer magnetic moment, $v$ is the skyrmion velocity vector, and $z$ is the unit vector normal to the conducting plane). This is a relevant issue for the development of skyrmion based devices because it can cause skyrmions to move toward the track edges, where they can become annihilated.

In antiferromagnets, a vanishing contribution from the skyrmion Hall effect, robustness against external magnetic fields, and a higher skyrmion racetrack speed reaching 900 m/s are key factors making antiferromagnetic skyrmions particularly promising for device development.[17] In contrast to ferromagnets where the Dzyaloshinskii-Moriya interaction (DMI) stabilizes skyrmions, the stability of antiferromagnetic skyrmions is governed by the antiferromagnetic Ruderman-Kittel-Kasuya-Yosida (RKKY) coupling, dipolar interactions, and current-induced Oersted fields.[14] There are a few reports exposing stable skyrmions in antiferromagnets which have yet to be fully explored for spintronics applications.[18-20] Magnetic systems with inherent non-coplanar spin textures, such as skyrmions, are characterized by a finite scalar spin chirality, $\chi_{ijk} = \boldsymbol{S}_i \cdot \boldsymbol{S}_j \times \boldsymbol{S}_k$, which was shown to lead to a pronounced topological Hall response already observed a in few antiferromagnetically ordered systems.[21-24] The observation of a topological Hall signal is usually accepted as evidence for spin chirality in occasions connected to topological spin textures. However, exposing their existence requires a combination of experimental techniques, such as Lorentz transmission electron microscopy (LTEM), X-ray transmission microscopy, or spin polarized scanning tunneling microscopy.

**Results/Discussion**

In this manuscript, we investigate the impact of Co doping on the magnetic domain structure of $Fe_3GaTe_2$ and its influence on the compound's transport and magnetic properties. Consistent with previous reports,[25] we find that ferromagnetism in $Fe_{3-x}Co_xGaTe_2$ is progressively suppressed with increasing Co content, culminating in a well-defined antiferromagnetic ground state when the Co fraction reaches approximately $x \sim (0.6 \pm 0.1)$. In contrast, Ni doping has been shown to rapidly suppress ferromagnetism at lower temperatures and stabilize a spin-glass state for Ni concentrations exceeding $x > 0.3$.[26] Below we show that Co-doping stabilizes a fragile, or metastable, antiferromagnetic state at a Co fraction dependent temperature 100 K $\leq T \leq$ 200 K which can be easily suppressed by relative small magnetic fields through a metamagnetic transition towards a spin polarized state. This metamagnetic transition leads to a clear anomalous Hall response as previously reported [26-31] and, surprisingly, to a topological Hall effect (THE) observed only at low temperatures. This points to the presence of topological spin textures as the moments become progressively polarized by the magnetic field. This motivated us to undertake a detailed magnetic force microscopy (MFM) study under an external magnetic field, which found a complex evolution of magnetic domains as the field drives this system from its AFM ground state to the spin polarized state. Attempts to image magnetic domains in $Fe_{3-x}Co_xGaTe_2$ using LTEM for compositions with $x \geq 0.6$ were unsuccessful, owing to the pronounced antiferromagnetic character of the material. In contrast, MFM finds that the observed THE appears only in a multidomain state characterized by the coexistence of up and down domains and, therefore, high domain-wall density, and a strong MFM contrast. All these conditions are met in a relatively narrow range of fields centered around $\mu_0 H \simeq \pm (4 \pm 0.5)$ T where the THE peaks. We conclude that around these field values, the field-driven reconfiguration of domains produces a network of chiral domain walls and skyrmion-like bubbles that are likely to carry a net topological charge and generate a large emergent magnetic field. ~~Remarkably,~~ such bubbles are observed already in the AFM phase, suggesting the existence of AFM skyrmions which are driven to ferromagnetic ones as the field increases.

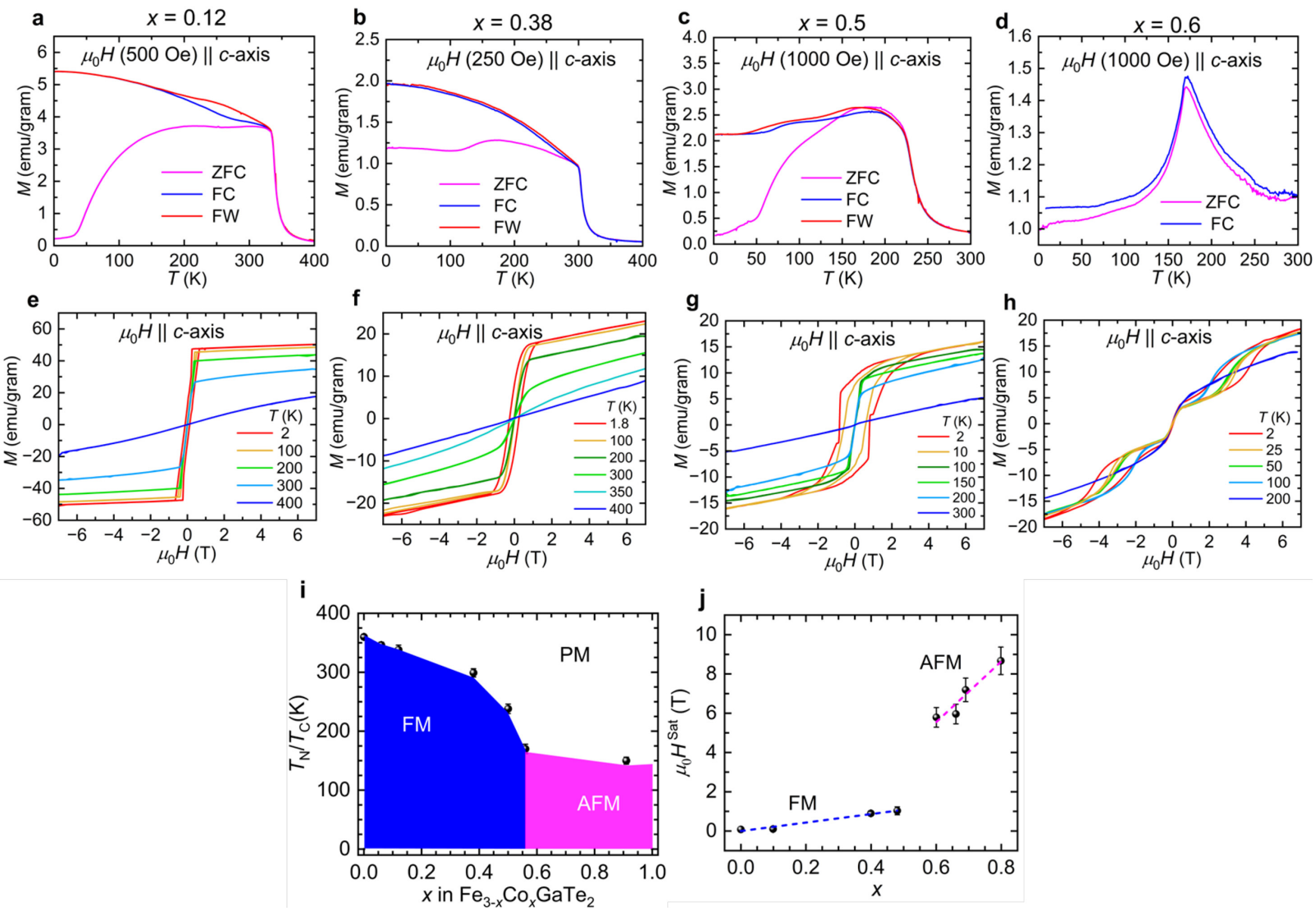

**Figure 1**. Magnetization as a function of temperature and magnetic field and resulting magnetic phase diagram of the $Fe_{3-x}Co_xGaTe_2$ system. (a-d) DC magnetization $M$ as a function of the temperature collected under zero field cooled (ZFC), field warmed (FW), and field cooled (FC) conditions for $Fe_{3-x}Co_xGaTe_2$ crystals having $x = 0.12$, 0.38, 0.5, and 0.6, according to X-ray refinement. (e-h) $M$ as a function of magnetic field $\mu_0 H$ along the $c$-axis of the crystals. Notice the increase in coercivity/hysteresis upon increasing the Co concentration, and the high metamagnetic transition at higher fields observed for $x = 0.6$, signaling the emergence of antiferromagnetism at higher Co concentrations. (i) Resulting phase diagram for the $Fe_{3-x}Co_xTe_2$ series, where blue colored area corresponds to ferromagnetic (FM) behavior, likely transitioning to ferrimagnetism at lower $T$s, and magenta one to antiferromagnetism (AFM). (j) Magnetic field values required to reach magnetization saturation $\mu_0 H^{Sat}$ for the $Fe_{3-x}Co_xGaTe_2$ system as a function of the Co concentration**.** The sharp increase in $\mu_0 H^{Sat}$ upon increasing the Co concentration beyond $x = 0.5$ indicates that i) the transition from FM to AFM is not continuous, and ii) that this system cannot characterized by the coexistence of phases or phase separation. Throughout this manuscript, indicated Co fractions were determined via elemental analysis.

We investigated Co-doped $Fe_3GaTe_2$ single crystals with varying Co concentrations, synthesized via chemical vapor transport (see, Methods in SI for details), using magnetotransport, magnetization, and magnetic force microscopy (MFM). Single-crystal X-ray diffraction[32-34] confirms that $Fe_{3-x}Co_xGaTe_2$ is isostructural to $Ni_3GeTe_2$, indicating the presence of intercalated Fe/Co sites.[35, 36] The intercalated (labelled as $Fe_3/Co_3$ site), situated within the vdW gap between the Te-layers, refines to an occupancy of ~10%. For the X-ray refinement Co was modeled as uniformly distributed among the three distinct Fe sites (details in the SI file). This system crystallizes in the same $P6_3/mmc$ space group of the parent $Fe_3GaTe_2$ compound (Figure S1 and Tables S1, S2).

As seen in Figure 1, we measured the magnetization of $Fe_{3-x}Co_xGaTe_2$, for single crystals with nominal concentrations $x = 0.12$, 0.38, 0.5, and 0.6 as a function of temperature $T$ for 1.8 K $\leq T \leq$ 300 K under external magnetic fields -7 T $\leq \mu_0 H \leq$ 7 T. These values of $x$, except for nominal $x = 0.5$ result from single crystal X-ray refinements (see, Table S1). $M$ as a function of $T$ was collected using the zero field-cooled (ZFC), field warmed (FW), and field-cooled (FC) protocols for fields applied along the $c$-axes of the crystals under $\mu_0 H$ = 500 Oe for $x = 0.2$, $\mu_0 H$ = 250 Oe for $x = 0.4$, $\mu_0 H$ = 1000 Oe for $x$ =0.5 and 0.7. The data clearly shows that the Curie temperature decreases as the Co concentration increases, in agreement with previous reports.[26-31] The middle point Curie and Néel temperatures are found to be $T_c$ = 337 K for $x = 0.12$, $T_c$ = 304 K for $x = 0.38$, $T_c$ = 229 K for $x = 0.5$, and $T_N$ = 172 K for $x = 0.6$ (Figures 1a to 1d). Similarly, the coercive fields are also found to increase as a function of Co concentration, as seen in Figures 1e to 1h, becoming antiferromagnetic when the Co concentration reaches $x > 0.5$. The resulting phase diagram (Figure 1i) and the dramatic increase in the magnetic field required to saturate the magnetization (Figure 1j), clearly indicate that for $x > 0.5$ the $Fe_{3-x}Co_xGaTe_2$ system becomes AFM. Figure 1 is provided here to indicate that the overall magnetic behavior of our crystals is akin to the one previously reported by other groups for this system. [26-31]

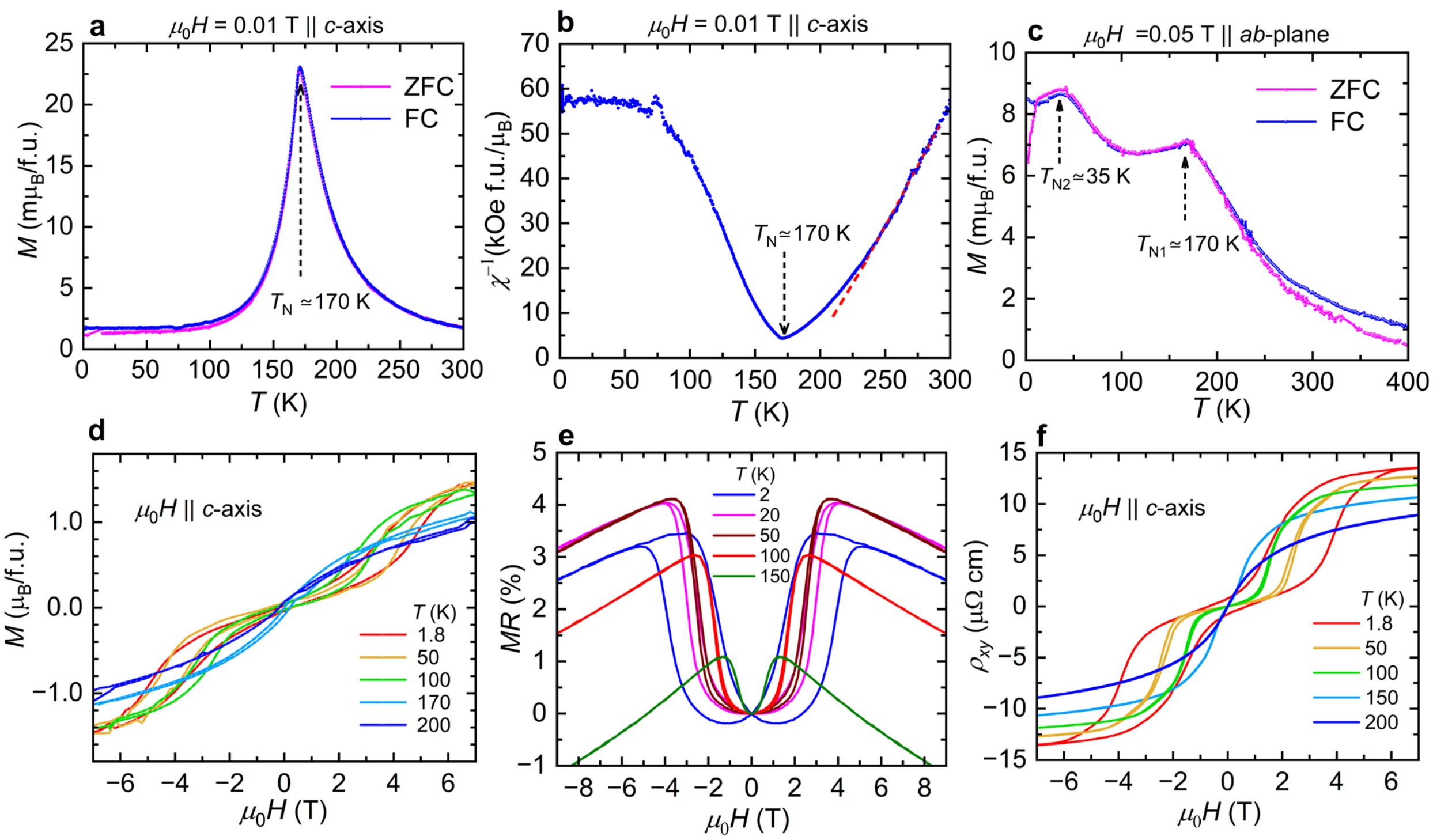

**Figure 2**. (a) $M$ as a function of the temperature $T$ for $Fe_{2.48}Co_{0.56}GaTe_2$ collected under both zero-field cooled (ZFC) and field-cooled (FC) conditions under an external magnetic field $\mu_0H = 0.01$ T along the $c$-axis of the crystal. Néel temperature is observed at $T_N \cong 170$ K. (b) Inverse magnetic susceptibility $\chi^{-1}$ as a function of $T$ collected under $\mu_0H = 0.01$ T along the c-axis of the crystal. Red dashed line is a Curie-Weiss fit above the $T_N$. (c) $M$ as a function of $T$ for a field $\mu_0H = 0.05$ T along the $ab$-plane of the crystal. A second transition is observed at $T_{N2} \sim 35$ K. (d) $M$ as a function of $\mu_0H$ along the $c$-axis. Hysteresis is observed in both the magnetization and concomitant anomalous Hall response. The saturation of $M$ occurs for fields $\mu_0H > 7$ T. (e) Magnetoresistivity MR = $(\rho_{xx}(\mu_0H) - \rho_0)/\rho_0$ where $\rho_0 = \rho_{xx}(\mu_0H = 0\text{ T})$ as a function of magnetic field $\mu_0H$ at different $T$s, showing a negative slope above a temperature dependent value of $\mu_0H$,

indicating spin polarization along the field. (f) Hall effect $\rho_{xy}$ as a function of $\mu_0 H$ applied along the *c*-axis of the crystal at different temperatures, revealing an anomalous Hall component that mimics the one observed in the magnetization. Size of hysteresis loops diminishes as $T$ increases.

Given that chiral or topological spin textures were reported for the pristine $Fe_3GaTe_2$ at room temperature and above,[5-9, 13] we chose to explore the possible presence of topologically non-trivial spin textures also in the Co-doped $Fe_3GaTe_2$ compounds via a detailed Hall effect study and concomitant analysis. For example, a seemingly topological Hall signal akin to the one claimed for pristine samples[37, 38] is also observed ferromagnetic $Fe_{2.89}Co_{0.12}GaTe_2$ (Figure S2). Consistent with the previous reports on these Co-doped compounds,[26-31] we find that there is a critical concentration of Co beyond which the system develops an AFM ground state. In our case, this critical value is found to be $x \sim 0.56$. Therefore, we choose this concentration to perform a systematic transport study that might unveil topology dominated transport once the AFM state is suppressed via an external field. $Fe_{2.48}Co_{0.56}GaTe_2$ single crystals were initially characterized by Energy dispersive X-ray spectroscopy (Figure S3). The crystal structure obtained from the refinement of the single crystal X-ray diffraction data is shown in Figure S4.

Figure 2a displays the magnetization $M$ of a $Fe_{2.48}Co_{0.56}GaTe_2$ crystal as a function of the temperature $T$ measured under an applied magnetic field $\mu_0 H = 100$ Oe applied along the *c*-axis. The zero field cooled (ZFC) and field cooled (FC) magnetization revealed no obvious hysteresis but showed a pronounced antiferromagnetic transition at $T_N \sim 170$ K. Figure 2b displays the inverse magnetic susceptibility as a function of $T$. One observes a nearly linear regime (red dashed line) above the antiferromagnetic transition and up to room temperature, implying Curie Weiss susceptibility. Figure 2c displays the magnetization of the crystal when an external magnetic field $\mu_0 H = 500$ Oe is applied along the *ab*-plane. In addition to the $T_{N1} \sim 170$ K, we observe an additional transition at a lower temperature $T_{N2}$ ~ 35 K, suggesting an additional spin reconfiguration occurring within the conducting planes of the material. Figure 2d displays the magnetization $M$ (in $\mu_B$ per f.u.) as a function of $\mu_0 H$, clearly indicating that $\rho_{xy}$ mimics $M$, hence the reason for the large coercive fields. To understand the electrical transport properties of $Fe_{2.48}Co_{0.56}GaTe_2$, we measured the resistivity $\rho_{xx}$ as a function of $T$ (Figure S5), as well as the magnetoresistivity $\rho_{xx}$ and the Hall resistivity $\rho_{xy}$ as functions of $\mu_0 H$. $\rho_{xx}(\mu_0 H)$ shows an initial quadratic dependence on $\mu_0 H$ that is followed by a sharp change in slope due to a metamagnetic transition (Figure 2e). The subsequent negative slope observed in the MR plot indicates that the mean free path of the charge carriers increases as the magnetic field increases due to the suppression of spin scattering in a spin polarized state. Here, the magnetoresistance is calculated as MR (%) = $[R_{xx}(\mu_0 H) - R_0(0)] \times 100 / R_0(0)$, with $R_{xx}(\mu_0 H)$ being the magnetoresistance measured via a 4-terminal configuration and $R_0(0)$ the resistance at 0 T. The residual resistivity at $T$ = 1.8 K was found to be 1.37 mΩ cm which is larger than the values measured in pristine $Fe_3GaTe_2$ which are in the order of 100 μΩ cm for the best crystals. This indicates that Co atoms act as impurities contributing significantly to carrier scattering. The residual resistance ratio (RRR) for this crystal is rather small RRR ≅ 1.34. Figure 2f shows the Hall resistivity for fields along the $c$-axis of the same crystal and for several $T$s. $\rho_{xy}$ reveals a large anomalous Hall component, i.e. $\rho_{xy}^A \propto M$, which is relatively uncommon for antiferromagnets. An anomalous Hall response in a centrosymmetric antiferromagnet usually requires i) non-collinear spin structures, ii) a combination of broken symmetries such as time-reversal and some lattice symmetry, iii) a finite Berry curvature due to spin–orbit coupling or chiral spin textures, or iv) multipole order.[39] This points to the unconventional nature of the AFM ground state in Co doped $Fe_3GaTe_2$. The large hysteresis loop in the anomalous Hall resistivity results from the formation and evolution of complex magnetic domains present in the system, as will be discussed below through the analysis of the magnetic force microscopy images.

We characterized the anomalous Hall conductivity (AHC) for the whole series of crystals belonging to the $Fe_{3-x}Co_xGaTe_2$ family having distinct Co content (Figure S6). When compared to the undoped $Fe_3GaTe_2$ compound, whose anomalous Hall conductivity reaches $\sigma_{xy}^{A}$ ~ 420 $\Omega^{-1}cm^{-1}$, the saturated value of the AHC for the Co doped/substituted samples is significantly smaller. Most likely this is due to strong scattering of charge carriers by impurities since the Co occupation in the Fe sites is assumed to be random. At $T$ = 1.8 K, the maximum value of the AHC (under $\mu_0H$ = 9 T) was found to be 52.27 $\Omega^{-1}cm^{-1}$, 36.99 $\Omega^{-1}cm^{-1}$, 58.49 $\Omega^{-1}cm^{-1}$, and 88.79 $\Omega^{-1}cm^{-1}$ for $x$ = 0.12, $x$ = 0.38, $x$ = 0.56, and $x$ = 0.91, respectively. As mentioned earlier, the existence of a complex magnetic domain structure in the antiferromagnetic state is evident from the larger coercive fields observed in the anomalous Hall conductivity (Figure S6c and S6d). To evaluate the possible existence of topologically non-trivial spin textures, we evaluated the topological Hall effect (THE) which could result from the aforementioned chiral particles.[40, 41] It is believed that a topological Hall signal constitutes experimental evidence for topological spin textures. THE is usually treated as an additive contribution to the overall Hall response and, therefore, it can be extracted by subtracting the conventional and AHE contributions:[42]

$$\rho_{xy} = \rho_{xy}^{\mathrm{N}} + \rho_{xy}^{\mathrm{A}} + \rho_{xy}^{\mathrm{T}} \quad (1)$$

$$\rho_{xy} = R_0 B + S_H M \rho_{xx}^2 + \rho_{xy}^{\mathrm{T}} \quad (2)$$

Therefore, the Hall resistivity of a material is usually analyzed as the sum of the conventional Hall component $\rho_{xy}^{\mathrm{N}}$ with $R_0$ being the Hall coefficient, and the anomalous Hall contribution $\rho_{xy}^{A}$ scaled as a function of the product of a power law of the magnetoresistivity $\rho_{xx}^2(\mu_0 H)$ and the magnetization $M$ through a proportionality constant, $S_H$. The topological Hall component $\rho_{xy}^{\mathrm{T}}$ can then be extracted from Eq. (2).

To obtain the topological contribution $\rho_{xy}^{\mathrm{T}}$ from a $Fe_{2.48}Co_{0.56}GaTe_2$ single crystal, we performed a scaling analysis (Figure 3) that clearly indicates that its Hall resistivity $\rho_{xy}$ depends linearly on the product of $M$ and $\rho_{xx}^2(\mu_0 H)$. A simple linear fit yields the anomalous Hall coefficient $S_H$. By subtracting the ordinary and anomalous Hall contributions from the raw $\rho_{xy}$, we extracted a clear $\rho_{xy}^{\mathrm{T}}$ signal that peaks at $\mu_0 H \sim$ - 4 T and 4 T upon increasing and decreasing the magnetic fields. In contrast, $\rho_{xy}^{\mathrm{T}}(\mu_0 H)$ is observed to increase linearly for -2 T ≤ $\mu_0 H$ ≤ 2 T. A similar analysis was applied to $Fe_{2.8}Co_{0.2}GaTe_2$ to estimate the $\rho_{xy}^{\mathrm{T}}$ contribution to the overall Hall response at lower Co concentrations when the $Fe_{3-x}Co_xGaTe_2$ series remains in the FM phase (see, Figure S2).

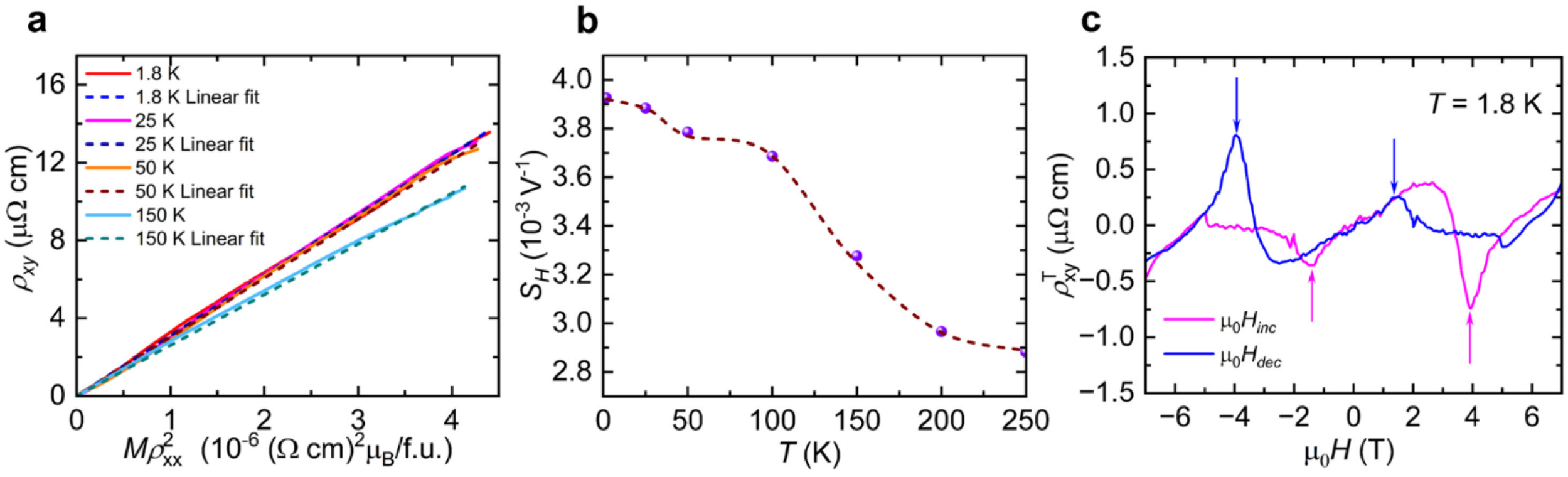

**Figure 3**. (a) Scaling of the anomalous Hall-effect as a function of $M\rho_{xx}^2$ where $M$ is the magnetization and $\rho_{xx}$ the magnetoresistivity for an AFM $Fe_{2.48}Co_{0.56}GaTe_2$ single crystal. (b) Anomalous Hall coefficient $S_H$, from the linear fits in panel (a) as a function of $T$. (c) Resulting topological Hall signal measured at $T$ = 1.8 K showing clear peaks at $\mu_0 H \sim -$ 4 T and 4 T. An almost linear increase of the topological Hall signal is observed for -2 T ≤ $\mu_0 H$ ≤ 2 T. As expected for a topological Hall signal, the observed peaks change sign upon reversing the sense of the magnetic field sweep.

Figure S8 displays the topological Hall-effect extracted from a second $Fe_{2.48}Co_{0.56}GaTe_2$ crystal collected from a second synthesis batch. This crystal displays a far more pronounced THE signal, likely resulting from a higher structural order or higher doping homogeneity, that is observable all the way up $T = 100$ K. This indicates that the second antiferromagnetic transition at $T_{N2}$ shown in Figure 2c, has little, to no effect, on THE response.

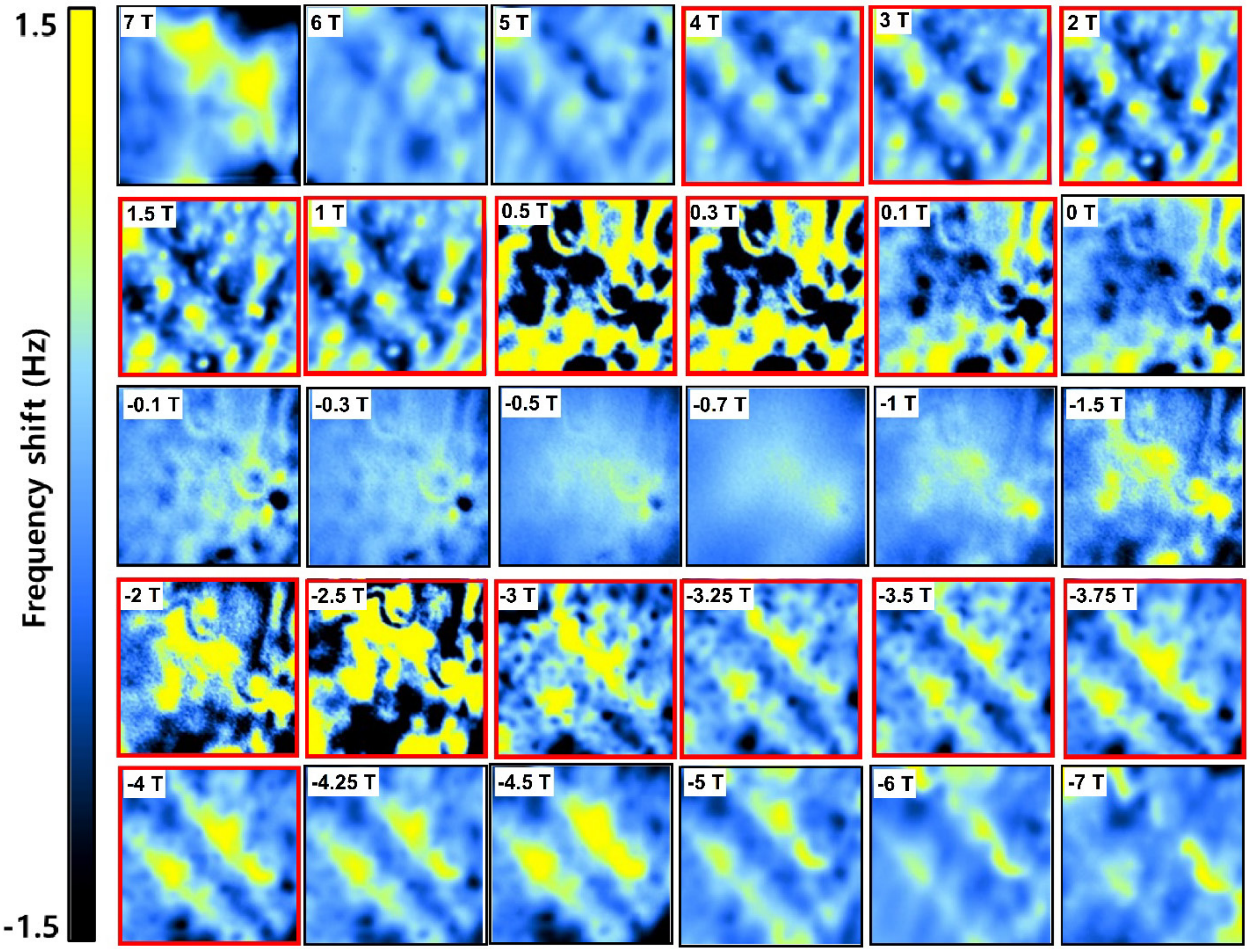

**Figure 4**. Magnetic force microscopy image for $Fe_{2.48}Co_{0.56}GaTe_2$ and for magnetic fields ranging from $-7$ T $\leq \mu_0 H \leq 7$ T with step-by-step increases in field intensity. Field values are indicated in each panel. Bubble-like features, possibly skyrmions, are clearly seen in those panels framed by red squares.

To investigate the unexpected observation of a topological Hall effect (THE) signal in an antiferromagnetic state, magnetic force microscopy (MFM) was employed to visualize the domain structure of $Fe_{3-x}Co_xGaTe_2$ as a function of magnetic field and temperature. In $Fe_3GaTe_2$, previous studies have reported the presence of spin textures, including Néel-type, Bloch-type, and hybrid skyrmions.[43,] MFM confirms that the $Fe_{3-x}Co_xGaTe_2$ series preserves the FM ground state at lower doping levels,[5, 6, 8, 9, 44] e.g., for nominal $x = 0.1$ (Figure S7) and $x = 0.12$. However, to understand the topological Hall signal observed at $T = 1.8$ K in Figure 3c and Figure S8, we performed a detailed investigation of the $Fe_{2.48}Co_{0.56}GaTe_2$ compound using a home-built MFM[45](Figure 4). MFM cannot resolve the texture of the spins since it detects only the out-of-plane component of the magnetization. However, MFM can still expose the domain structure or, as we show below, the possible presence of more complex and topological spin textures, such as skyrmions.

To this effect, we examined a possible correlation between the topological Hall resistance and the evolution of the domain structure captured by MFM images. Therefore, the MFM measurements were performed following the same magnetic-field sequence previously used for the Hall measurements. The scan conditions are as follows: scan size of 30 × 30 μm² with a pixel size of 128 × 128, temperature $T = 4.2$ K, and lift height $h = 200$ nm. First, the sample was saturated by applying a magnetic field $\mu_0 H = +7$ T, which was gradually decreased to $\mu_0 H = -7$ T. In the MFM images, a positive frequency shift (yellow contrast) corresponds to downward magnetization, whereas a negative frequency shift (dark blue) corresponds to upward magnetization.

As shown in Figure 4, starting from the positively saturated state at $\mu_0 H = +7$ T, the contrast is relatively weak and spatially smooth, indicating an almost uniform out-of-plane magnetization. As the field is reduced to $\mu_0 H = +3$ T and +2 T, the MFM images develop bright bubble domains, and these field points already belong to the highlighted texture-evolution window, while the characteristic lateral length scale of the contrast remains nearly unchanged. Upon further decreasing the field to +1.5 T and especially to +1 ~ +0.3 T, the images evolve into a strongly modulated multidomain state with dense stripe-, bubble-, and labyrinth-like features, signaling the formation of a dense multidomain texture. When the field is swept through zero and into the negative regime, the contrast becomes again relatively homogeneous at $\mu_0 H = -0.7$ T, indicating a transiently more uniform magnetization. However, upon increasing negative fields, a complex domain pattern reappears: the strongest texture modulation is observed in the field range from about −3 to −4 T, where a dense stripe/bubble-like texture is clearly visible and remains highlighted. By contrast, the very strong black/yellow labyrinthine-like patterns around −2 to −2.5 T appear even more disordered, indicating that maximal visual contrast does not necessarily coincide with the field window of the largest topological Hall signal. Throughout the entire field sweep from +7 T to −7 T, the typical domain size varies only moderately, whereas the relative fraction of bright and dark areas and the overall contrast change markedly. This behavior differs from that of a simple ferromagnet, where domains usually coarsen and merge into a single-domain state as the field is increased, and instead suggests that the magnetization is largely redistributed within a quasi-fixed domain mosaic. The topological Hall resistivity extracted in the same sweep protocol shown in Figure 3c shows a large positive peak around $\mu_0 H \simeq -4$ T on the negative-field branch and a shallow maximum on the positive-field side. The strongest THE appears in the field range where the MFM images reveal a highly modulated multidomain pattern, indicating that non-coplanar spin textures are stabilized within this regime.

To further clarify the pronounced asymmetry of the topological Hall effect seen in Fig. 3c, we performed additional MFM measurements for both decreasing- and increasing-field sweeps, as shown in Supplementary Figure S10. The new data reveals that the magnetic textures evolve along markedly different pathways depending on sweep direction, rather than retracing the same domain configurations upon field reversal. Consistent with the field windows highlighted in Fig. 4, the decreasing-field sweep stabilizes strongly modulated stripe- and bubble-like textures near $\mu_0 H \approx$ +1.5 to +1 T and again near $\mu_0 H \approx -3$ to −3.25 T, the latter coinciding with the large positive $\rho_{xy}^{\mathrm{T}}$ peak on the $\mu_0 H_{dec}$ branch. In contrast, during the increasing-field sweep, the strongest multidomain contrast appears in different field windows, most notably around $\mu_0 H \approx -1.25$ to −0.75 T and $\mu_0 H \approx +3.25$ to +3 T, consistent with the shifted THE anomaly on the $\mu_0 H_{inc}$ branch. These observations indicate that the asymmetry of $\rho_{xy}^{\mathrm{T}}$ is governed primarily by hysteretic, sweep-history-dependent evolution of the real-space magnetic texture across the metamagnetic transition, rather than by a simple intrinsic field asymmetry.

To estimate the intrinsic size of the bubble-like textures observed in Fig. 4, we compared them with superconducting vortices in a Nb film measured using the same MFM tip in the same cooldown sequence (Supplementary Fig. S12). Under our imaging conditions, Nb vortices with an intrinsic magnetic length scale of order 100 nm appear with an apparent diameter of about 2 μm. The comparable apparent size of the Co-doped $Fe_3GaTe_2$ bubbles therefore indicates that their intrinsic diameter is also likely much smaller than the raw MFM width, on the order of a few hundred nanometers or less.

Next, we quantify these observations by extracting several scalar metrics that characterize domain morphology from the complete set of MFM images and correlating them with the measured topological Hall resistivity $\rho_{xy}^{\mathrm{T}}$. From each MFM image we extracted three scalar quantities: 1) the domain-wall length $L_{\mathrm{DW}}$ or the total length of domain walls, obtained from Canny edge detection[46] after a mild Gaussian smoothing, 2) up-domain area fraction $f_{\uparrow}$ or the fraction of the image area where the MFM frequency shift exceeds an Otsu threshold[47], corresponding to domains with positive out-of-plane magnetization, and 3) intensity variance Var(I) or the variance of the MFM frequency-shift signal, which reflects the overall contrast and complexity of the domain pattern. These three metrics and the corresponding $\rho_{xy}^{\mathrm{T}}$ are plotted as a function of $\mu_0 H$ in Figures. 5a–5c. Figure 5a shows the field dependence of the domain-wall length $L_{\mathrm{DW}}$ (blue, left axis) together with the topological Hall resistivity $\rho_{xy}^{\mathrm{T}}$ (red, right axis). At large negative and positive fields ($|\mu_0 H| \gtrsim 5$ T), $L_{\mathrm{DW}}$ assumes relatively small, on the order of a few hundred pixels, while $\rho_{xy}^{\mathrm{T}}$ is nearly zero. In this regime, the MFM images display weak and comparatively uniform contrast, consistent with a nearly field-polarized state in which extended multidomain textures are absent**.** This indicates that non-coplanar spin textures do not significantly develop in the high-field regime.

Starting from the saturation observed at $\mu_0 H$ = +6 T and upon decreasing the field, $L_{\mathrm{DW}}$ increases gradually as the system approaches the intermediate-field region. On the positive-field side, a broad enhancement of $L_{\mathrm{DW}}$ is observed between approximately +1 and +3 T, where $\rho_{xy}^{\mathrm{T}}$ also exhibits a weaker positive hump**.** This indicates that a multidomain state with a finite density of domain walls already forms on the positive-field branch, but the associated topological Hall response remains modest.

As the field is decreased through $\mu_0 H$ = 0 T and into negative values, $L_{\mathrm{DW}}$ again increases sharply, with pronounced maxima appearing in the range from about -4.5 T to -2 T. In the same field interval, $\rho_{xy}^{\mathrm{T}}$ rises strongly and reaches its largest positive value near $\mu_0 H \approx -4.5$ to $-4$ T. Notably, the maximum of $\rho_{xy}^{\mathrm{T}}$ does not coincide exactly with the absolute maximum of $\mathrm{L_{DW}}$: the domain-wall length remains high and even shows comparable or larger values at somewhat higher fields, whereas the topological Hall response is already reduced. This demonstrates that a large domain-wall density is a necessary condition for the emergence of THE, but not a sufficient one.

The overall behavior therefore reveals a clear but non-monotonic correlation between $L_{DW}$ and $\rho_{xy}^{\mathrm{T}}$ . A finite and extended network of domain walls is required for the appearance of a sizeable topological Hall signal, but the magnitude of $\rho_{xy}^{\mathrm{T}}$ is not determined solely by the total amount of domain walls. Rather, it also depends on the detailed topology and chirality of the domain-wall network. In particular, $\rho_{xy}^{\mathrm{T}}$ is maximized in an intermediate field window where the domain-wall structure is already well developed but not yet in its most disordered state, suggesting that the real-space texture is optimized for generating a net emergent magnetic flux. This is consistent with the formation of skyrmion-like bubbles, such as the circular domains highlighted in the red-framed panels of Fig. 4. At lower fields, by contrast, the domain pattern may become so intricate that textures with opposite topological charge coexist and partially cancel each other in the macroscopic transport signal. Overall, Fig. 5a establishes that the emergence of a large topological Hall response is intimately tied to the formation of a dense, field-dependent domain-wall network, and that the maximum in $\rho_{xy}^{\mathrm{T}}$ is realized in a narrow field window on the negative-field branch where the real-space topology of the spin texture is most favorable.

Figure 5b correlates the sign-corrected up-domain area fraction $f_{\uparrow}$ (blue) with $\rho_{xy}^{\mathrm{T}}$ (red). Because the MFM contrast primarily reflects the out-of-plane component of the magnetization, the domain fraction provides a measure of how strongly one magnetic polarity dominates the multidomain state. In the present analysis, the raw bright-domain fraction was corrected for the field polarity so that the plotted quantity consistently represents the field-aligned domain fraction on both the positive- and negative-field sides. In this way, Fig. 5b allows a more physically meaningful comparison between the domain population imbalance and the topological Hall response over the entire decreasing-field sweep. The most important observation is that the largest topological Hall signal occurs not in the fully demagnetized state ($f_\uparrow \approx 0.5$), nor in the saturated state ($f_\uparrow \rightarrow 0$ or 1), but rather in a partially magnetized multidomain regime.

Around the negative-field THE maximum at $\mu_0 H \simeq -4.5$ to 4 T, the sign-corrected domain fraction lies in the range $f_\uparrow \simeq 0.55 - 0.65$. In this window, the sample contains both up and down domains in comparable amounts, but one polarity is already favored, so that the system retains a net magnetization. Yet, it retains a net magnetization. This indicates that the THE requires a compromise between two conditions: 1) the coexistence of up and down domains to allow for non-collinear and non-coplanar spin arrangements at domain boundaries and within complex textures; 2) a finite net magnetization that selects a preferred sense of rotation and prevents complete cancellation of the emergent fields from textures of opposite chirality. For fields where $f_\uparrow$ is very close to 0 or 1, the sample becomes essentially single-domain, domain walls vanish, and $\rho_{xy}^{\mathrm{T}}$ tends to zero. Conversely, in a perfectly demagnetized state with comparable fractions of the two domain polarities, one might expect equal populations of textures with positive and negative topological charges, again strongly reducing the net THE.

The MFM data support this interpretation: the topological Hall resistivity is enhanced only within a narrow range, $0.5 \lesssim f_\uparrow \lesssim 0.65$, where both up and down domains coexist while a clear global magnetization direction is maintained. The fact that fields with similar $f_\uparrow$ values—for example, around $\mu_0 H \cong -4$ T and +1 to +2 T— are encountered at different stages of the decreasing-field sweep yet exhibit markedly different $\rho_{xy}^{\mathrm{T}}$ underscores that the average magnetization alone does not determine the topological Hall response.

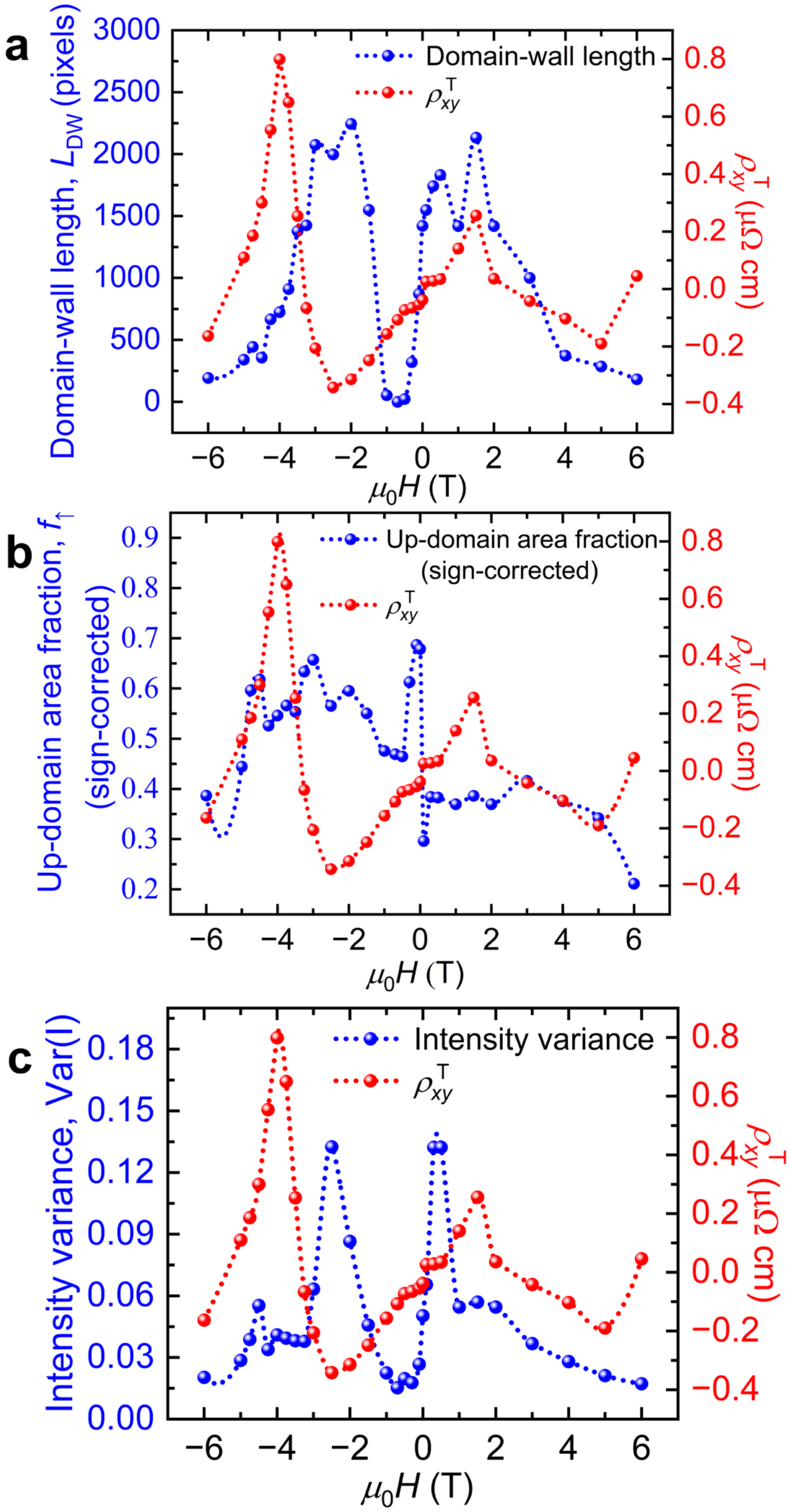

**Figure 5**. Evolution of the MFM domain metrics and topological Hall response collected upon decreasing the external field $\mu_0 H$ from +6 T to −6 T from a $Fe_{2.48}Co_{0.56}GaTe_2$ single crystal. (a) Domain-wall length

$L_{DW}$ (blue, left axis) and topological Hall resistivity $\rho_{xy}^{T}$ (red, right axis). (b) Up-domain area fraction $f_{\uparrow}$ (blue) and $\rho_{xy}^{T}$ (red). (c) Intensity variance Var(I) (blue) and $\rho_{xy}^{T}$ (red).

Instead, it acts as a necessary but not sufficient condition; the detailed domain morphology and its chirality, as captured better by the other metrics and techniques, must also be considered.

Thus, Fig. 5b indicates that the THE is maximized not simply by maximizing the field-aligned domain fraction, but by stabilizing a partially aligned multidomain state in which topologically nontrivial, non-coplanar textures can develop most effectively. Figure 5c displays the field dependence of the variance of the MFM frequency-shift signal, Var(I), together with $\rho_{xy}^{T}$ . as a function of field. This quantity reflects the overall contrast and spatial heterogeneity of the MFM images: it increases when bright and dark regions coexist strongly and when the domain pattern becomes more modulated. Like $L_{DW}$, Var(I) is small in the high-field saturated states and increases as the system enters the multidomain regime upon decreasing $|\mu_0 H|$ from +7 T toward −7 T. On the negative-field side, Var(I) rises substantially in the same field region where $\rho_{xy}^{T}$ also becomes enhanced, indicating that domain nucleation and growth enhance the overall MFM contrast In particular, $\rho_{xy}^{T}$ reaches its largest positive value near $\mu_0 H \approx -4.5$ to −4 T, where Var(I) is already significantly enhanced but has not yet reached its absolute maximum. This concurrence reinforces the conclusion that strongly modulated, labyrinth- or bubble-like domain configurations are favorable for stabilizing non-coplanar spin textures responsible for the THE.

At the same time, the maximum of Var(I) does not coincide exactly with the maximum of $\rho_{xy}^{T}$ . Rather, Var(I) shows its strongest peaks near $\mu_0 H \approx -2.5$ T and again close to 0 T, where $\rho_{xy}^{T}$ has already begun to decrease.  A similar tendency-though with smaller amplitude-is seen on the positive-field side around $\mu_0 H \sim +1$ to +2 T, which is encountered earlier in the decreasing-field sweep. Such behavior again highlights a stronger MFM contrast does not automatically imply a larger topological Hall response. At very high contrast, the domain structure may become excessively disordered, hosting a mixture of textures with opposite topological charges whose contributions to the emergent field partially cancel in the macroscopic transport signal. Thus, the intensity variance serves as another indicator that the THE is optimized in an intermediate regime: the domain pattern must be sufficiently complex and modulated to host non-coplanar structures, but not so disordered that the net emergent field averages out. In this sense, Fig. 5c complements Figs. 5a and 5b by showing that the largest topological Hall response is realized not at the point of maximal contrast, but in the field window where the contrast is already strongly enhanced while the texture remains sufficiently organized to sustain a nonzero net chirality.

Additional insight is obtained by directly inspecting the sequence of MFM images in Figure 4, which were acquired during a monotonic field sweep from +6 T to −6 T. In a conventional ferromagnet governed primarily by simple domain-wall motion, one expects the domain size to increase and domains to gradually merge into a nearly single-domain state as $|\mu_0 H|$ is increased. Our MFM images instead show a distinctly different evolution. In contrast, our images show a rather different evolution. Starting from the positively saturated state at +6 T, the contrast is already weakly modulated rather than completely featureless. As the field is decreased towards +2 T and +1.5 T, bright bubble domains emerge, while the typical lateral scale of the contrast remains approximately unchanged, while the relative brightness of different regions evolves. Upon approaching $\mu_0 H \simeq +0.5$ T, the contrast suddenly develops a labyrinthine-like pattern with strong alternation between bright and dark areas. A similar re-entrant behavior is observed on the negative-field side: the image becomes comparatively uniform again near -2 T, whereas a dense network of stripes and bubble-like domains reappears for $\mu_0 H$ = -2.5 T to $\mu_0 H = -3.75$ T. Overall, the characteristic domain size changes much less dramatically than in a simple ferromagnet; instead, the magnetization within a given domain pattern seems to be continuously redistributed as the field is swept.

Such behavior is more reminiscent of ferrimagnetic or antiferromagnetic systems, in which the net magnetization results from the partial cancellation of two or more antiparallel sublattices. In that case, the microscopic spin texture (the Néel order) can undergo substantial reorientation while the macroscopic domain pattern is largely pinned by structural inhomogeneities. Within this picture, the evolution of the MFM contrast in Figure 4 can be interpreted as a change of the effective magnetization inside a quasi-fixed domain mosaic, rather than a simple growth and shrinkage of domains as in a conventional ferromagnet.

The correlation between this "ferri/antiferromagnetic"-like domain evolution and the topological Hall response raise the question of whether the underlying spin textures could be related to ferrimagnetic or even antiferromagnetic skyrmions. In a purely ferromagnetic skyrmion lattice, the skyrmion diameter usually shrinks monotonically with increasing field and collapses at a critical field, whereas in compensated ferrimagnets or antiferromagnets, the skyrmion core size can remain nearly field independent and the net magnetization can change mainly through the internal canting of the sublattices. Our MFM observations align more closely with the latter scenario: the lateral scale of the contrast associated with the labyrinthine or bubble-like patterns, remains relatively stable under varying magnetic fields, while both the average magnetization (quantified by $f_{\uparrow}$) and the contrast amplitude exhibit pronounced changes.

From the transport side, the large topological Hall resistivity around $\mu_0 H \simeq -3.5$ T appears in a field range where the net magnetization is only partial, yet the domain-wall density and intensity variance are both high. This combination—weak or moderate macroscopic magnetization with a strong THE—is naturally expected if the real-space Berry curvature originates from non-coplanar spin textures on (partially) compensated sublattices. In a ferrimagnetic skyrmion scenario, two antiferromagnetically coupled sublattices form skyrmions with opposite core magnetizations but unequal moments, so that the emergent fields from the two sublattices add, while the uniform magnetization partially cancels. MFM, which is sensitive mainly to the uncompensated stray field, would then detect a relatively weak albeit still finite contrast, consistent with the rather soft contrast in some of the images where $\rho_{xy}^{\mathrm{T}}$ is already sizable.

At the same time, our present measurements do not allow us to unambiguously identify antiferromagnetic skyrmions in the strict sense of perfectly compensated Néel textures. As already mentioned, MFM is insensitive to the internal spin structure of each magnetic sublattice and essentially probes only the net out-of-plane moment. A truly compensated antiferromagnetic skyrmion lattice would generate almost no stray field and therefore be nearly invisible in MFM. Even if such a texture were present, it could still contribute to a substantial topological Hall effect through its emergent field once a finite uncompensated component is induced by the external field. The clear contrast and hysteresis seen in our images indicate that a sizable uncompensated component remains, suggesting that the system is better described as a ferrimagnet (or a weak ferromagnet) hosting chiral, skyrmion-like textures rather than a purely antiferromagnetic skyrmion lattice.

We therefore conclude that the unusual field evolution of the domain pattern — in which the domain mosaic is relatively robust while the internal magnetization and contrast change strongly as the field is swept from $\mu_0 H = +7$ T to $\mu_0 H = -7$ T — is qualitatively consistent with the presence of ferrimagnetic skyrmion-like textures whose real-space topology is most pronounced on the negative-field branch. However, demonstrating bona fide antiferromagnetic skyrmions would require element-specific or sublattice-sensitive probes, such as resonant X-ray microscopy or spin-polarized scanning tunneling microscopy, which can directly resolve the Néel order on each sublattice. Within the present data set, the safest interpretation is that the large topological Hall signal arises from non-coplanar spin textures in a partially compensated, ferrimagnetic-like domain state, whose field evolution differs markedly from that of a simple ferromagnet.

**Conclusions**

In summary, we synthesized and characterized the $Fe_{3-x}Co_xGaTe_2$ series confirming previous reports on the crystal structure and the overall evolution of the electrical transport properties and magnetic ground states upon increasing Co content. The Curie temperature of the ferromagnetic state indeed decreases progressively, while the magnetic coercive fields increase as a function of $x$. For Co fractions beyond $x \approx 0.6$, doping drives the ferrimagnetic ground state of the pristine compound into an antiferromagnetic one as clearly indicated by the magnetization data and the inability of LTEM to visualize the domain structure seen via MFM. Ferrimagnetism, as reported for the parent compound, would not preclude the observation of spin textures via LTEM. Therefore, we conclude that the topological Hall signal measured by us in several crystals ought to result from chiral spin textures, possibly AF skyrmions, which become progressively polarized as the external field is increased.

~~Remarkably,~~ we observe clear evidence for a topological Hall effect at low temperatures when the AFM state is suppressed by an external magnetic field and transformed into a spin-polarized state through a metamagnetic transition. This sizeable THE appears only within an intermediate multidomain regime, where both domain polarities coexist, the domain-wall density is high, and the MFM contrast is strongly enhanced. In the revised quantitative analysis, these conditions are met in a relatively narrow field window centered around $\mu_0 H$ = -4 T on the decreasing-field branch. Taken together, our results demonstrate that THE in $Fe_{2.48}Co_{0.56}GaTe_2$ is intimately connected to the field-induced evolution of the magnetic domain structure.

As for the origin of the topological spin textures, according to the X-ray data in the SI file, $Fe_{1-x}Co_xGaTe_2$ crystallizes in the same centrosymmetric $P6_3mmc$ space group as the parent undoped compound. In Ref. [13] we already exposed a local lack of inversion symmetry in $Fe_3GaTe_2$ due to structural disorder at the Fe2 sites leading to interstitial Fe3 sites. We proposed that local lack of inversion symmetry, and a certain level of magnetic frustration resulting from structural disorder and competing ferromagnetic and antiferromagnetic interactions are the essential ingredients leading to skyrmion formation in these compounds. Notice that we also reported $Fe_3GaTe_2$ to display a ferrimagnetic instead of a ferromagnetic ground state[13]. The overall experimental evidence indicates that Co doping at the Fe sites stabilizes AF. But ferromagnetic exchange interactions remain relevant as indicated by the relatively small fields ($\mu_0 H \cong 4$ T – 6 T), relative to the Néel temperature $T_{N1} \cong 170$ K – 200 K, required to suppress the AF ground state and to polarize the moments of $Fe_{2.48}Co_{0.56}GaTe_2$. Hence, topological spin textures in this compound result from disorder induced local lack of inversion symmetry, and competing magnetic interactions.

Given the asymmetry between increasing and decreasing magnetic field sweeps, where a putative THE signal, or a peak, is observed *mainly* upon increasing the field precisely as reported for skyrmionics systems[48], one cannot attribute its observation to domain coexistence. Hysteresis, or domain coexistence, is observed in both traces, or upon increasing and decreasing the field across the phase boundary between AF and the spin polarized state. If domain coexistence, or boundary, was the origin of this peak one should clearly observe it on both traces.

In any case, the combination of quantitative MFM analysis and transport measurements provides direct experimental evidence indicating that the large THE originates from non-coplanar spin textures emerging in a partially magnetized multidomain state with a dense network of chiral domain walls, which seems to lead to skyrmion formation. Given that such multi-domain structures are already observed in the antiferromagnetic state prior to its magnetic field induced suppression, we propose that antiferromagnetic skyrmions are present in this family of compounds, and wonder if distinct doping strategies could bring such objects up to room temperature.

We point out that the presence of a topological Hall response in the parent $Fe_3GaTe_2$ compound remains controversial with different groups reporting distinct responses[37, 38]. Our previous study on $Fe_3GaTe_2$ found that the magnitude, and shape of THE signal depends on sample quality, scaling with anomalous Hall resistivity instead of the anomalous Hall conductivity, or the conductivity of the crystal[13]. This indicates that the putative THE signal in $Fe_3GaTe_2$ scales with scattering or disorder and therefore cannot be attributed to skyrmion density. Therefore, we attributed it to an anomaly in the Hall resistivity associated with the suppression of ferrimagnetism. In contrast, THE signals shown in Figures 3 and S8 are asymmetric; peaks are only observed upon increasing the magnitude of the magnetic field. This behavior cannot be reconciled with a phase transition which should observed upon increasing and decreasing magnetic field. Therefore, the magnitude of the THE signal reported here cannot be compared to the ones reported for the $Fe_3GaTe_2$ compound. It is nevertheless surprising that one detects a clear THE signal in an AF compound, while its presence in the parent ferrimagnetic compound remains a matter of debate despite multiple reports of skyrmions[6, 9, 44].

It is worth noting that structural disorder, such as atomic intercalation, can break mirror and inversion symmetries, thereby enabling the Dzyaloshinskii–Moriya interaction, which favors skyrmion formation.[13, 49] This mechanism may also account for the observation of an anomalous Hall response in an antiferromagnet that is, in principle, centrosymmetric. If confirmed by other techniques, Co doped $Fe_3GaTe_2$ might become a rare example of ~~become to the first~~ single-crystalline compound naturally displaying antiferromagnetic skyrmions, usually observed in synthetic antiferromagnet multilayers. Finally, during the review process of our manuscript we became aware of Ref. [50] whose authors claim to observe chiral spin textures and antiferromagnetic skyrmions in Co doped $Fe_3GaTe_2$ via spin-polarized low-energy electron microscopy.

## METHODS/EXPERIMENTAL

### Crystal growth.

Single crystals of Co doped $Fe_{3-x}Co_xGaTe_2$ were synthesized through a chemical vapor transport method. High purity Fe and Co powders, Ga lumps, and Te lumps were mixed in the molar ratio (3-$x$):$x$:1:2 and sealed in an evacuated quartz tube. The mixture was pre-reacted at 1000 °C for 24 hours. The resulting product was finely ground in an agate mortar. The pre-reacted powder was sealed in a vacuum quartz tube with a small amount of iodine (5 mg/cm$^3$). The tube was then placed in a two-zone horizontal tube furnace with a temperature gradient of 750 - 720 °C for two weeks. The tube was quenched into ice-water. The resulting crystals were washed with acetone and isopropanol.

### Single crystal X-ray diffraction.

Single-crystal X diffraction data were obtained at room temperature using small crystals of $Fe_{2-x}Co_xGaTe_2$. Data was collected using a Bruker D8 Quest Kappa single-crystal X-ray diffractometer equipped with an Incoatec IμS microfocus source (Mo Kα radiation, λ = 0.71073 Å) and a PHOTON III CPAD area detector. The raw frames were integrated with Bruker SAINT, and the intensities were corrected for absorption with a multi-scan method in SADABS 2016/2.[32] The intrinsic phasing method in SHELXT was used to generate preliminary crystallographic models[33], which were finalized with least squares refinements in SHELXL2019.[34] The details for the crystallographic data, data collection, and the refinement parameters are outlined in Table S1 below. The crystallographic structure of $Fe_{3-x}Co_xGaTe_2$ (for $x$ = 0.12, 0.38, 0.56, 0.6, 0.91, and 1.05) consists of five atomic sites as outlined in Table S2. There is a contraction of the lattice parameters (~0.025 Å along $a$, ~0.05 Å along $c$, and ~3 Å$^3$ in volume) with increasing Co concentration (depicted in Figure S1).

**Electrical transport and magnetization measurements.**

The electrical transport measurements were performed in a Physical Property Measurement System (Quantum Design PPMS). The resistivity of the crystals was measured using a four-probe method. Magnetoresistance and Hall resistance were measured from $T$ = 300 K to 1.8 K by sweeping the magnetic field from $\mu_0 H$ = -9 T to $\mu_0 H$ = 9 T. DC magnetization $M$ was collected in a Magnetic Property Measurement System (Quantum Design MPMS-XL). Field cooled (FC) and zero field cooled (ZFC) measurements were measured from $T$ = 300 K to $T$ = 1.8 K. The magnetization of the sample was measured as a function of external magnetic field for values ranging from $\mu_0 H$ = -7 T to $\mu_0 H$ = 7 T.

**Magnetic force microscopy.**

The individual magnetic domains of the $Fe_{2.48}Co_{0.56}GaTe_2$ crystals were imaged using a home-built Magnetic Force Microscopy (MFM) at $T$ = 4.2 K. While Figure 4 describes the texture evolution during a single monotonic field sweep from +7 T to -7 T, Figure S10 presents the extended MFM datasets covering both sweep directions: from +7 T to -7 T and the reverse path from -7 T to +7 T. The system is completely polarized under 7 T. Images were taken in certain field intervals upon varying the field. To clarify how the real-space magnetic textures give rise to the topological Hall effect (THE) in this material, we quantitatively analyzed MFM images taken as a function of out-of-plane magnetic field $\mu_0 H$ at $T$ = 4.2 K and compared them with the topological Hall resistivity $\rho_{xy}^{\mathrm{T}}$ ($\mu_0 H$) measured at 1.8 K.

**Quantitative analysis of MFM domain morphology.**

To quantify the evolution of the magnetic domain structure as a function of magnetic field, we analyzed each MFM image using a common image-processing pipeline. Figure S9 illustrates the procedure for a representative field.

1) Normalization and smoothing. The raw MFM frequency-shift image is converted to a grayscale matrix and normalized to the range [0,1]. A mild Gaussian filter (typical kernel σ = 1–2 pixels) is applied to suppress high-frequency noise without altering the overall domain morphology.

2) Binary segmentation and up-domain fraction $f_{\uparrow}$. An intensity threshold is determined by Otsu's method, which maximizes the inter-class variance between "bright" and "dark" pixels. Pixels above (below) the threshold are assigned to up (down) domains, yielding a binary map [Fig. 5(b)]. The up-domain area fraction is then computed as $f_{\uparrow} = N_{\mathrm{up}} /(N_{\mathrm{up}} + N_{\mathrm{down}})$, where $N_{\mathrm{up}}$ and $N_{\mathrm{down}}$ are the number of pixels in each class.

3) Domain-wall extraction and length $L_{\mathrm{DW}}$. We apply a Canny edge detector to the smoothed grayscale image to extract the boundaries between bright and dark regions [Fig. 5(c)]. Small speckles consisting of fewer than $N_{\mathrm{min}}$ connected pixels (typically $N_{\mathrm{min}}$ = 10) are removed to avoid counting noise as domain walls. The remaining connected edge pixels form extended domain walls. The total domain wall length is then proportional to the number of edge pixels: $L_{\mathrm{DW}}$ = a × $N_{edge}$, where a denotes the pixel size in real space. In the main text we plot $N_{edge}$ (in pixels) as a convenient measure of $L_{DW}$

4) Intensity variance. Finally, we compute the variance of the normalized intensity over the entire image, $Var(I) = \langle I^2 \rangle - \langle I \rangle^2$, which quantifies the overall domain contrast and mixture of bright and dark regions. Each of these metrics is evaluated for all MFM images and plotted as a function of magnetic field in Fig. 5(a)–(c) of the main text.

**Sweep-direction dependence of the MFM textures and topological Hall response.**

To further examine the branch asymmetry of the topological Hall effect, we performed additional MFM imaging for both decreasing- and increasing-field sweeps and compared the corresponding domain evolution with the field dependence of $\rho_{xy}^{T}$ . The results are summarized in Supplementary Figure S10. In the corrected labeling of Fig. 3c, the decreasing-field sweep is denoted by $\mu_0 H_{dec}$ (blue), whereas the increasing-field sweep is denoted by $\mu_0 H_{inc}$ (magenta).

The MFM images clearly show that the magnetic textures do not retrace the same sequence upon reversing the field sweep direction. During the decreasing-field sweep from +7 T to −7 T, the system evolves from a nearly field-polarized state into a strongly modulated multidomain texture in two characteristic field windows highlighted by the red boxes: first around $\mu_0 H \approx +1.5$ to +1 T, and second around $\mu_0 H \approx -3$ to -3.25 T. The latter field range coincides closely with the strongest positive peak of $\rho_{xy}^{T}$ observed on the $\mu_0 H_{dec}$ branch. These images show a dense stripe- and bubble-like morphology, suggesting that this branch stabilizes a particularly favorable non-coplanar texture for generating a large emergent field.

By contrast, during the increasing-field sweep from −7 T to +7 T, the magnetic texture evolves through a distinctly different sequence of states. The strongest multidomain contrast is observed in the red-boxed field windows around $\mu_0 H \approx -1.25$ to $-0.75$ T and $\mu_0 H \approx +3$ to $+3.25$ T, rather than in the same field ranges emphasized during the decreasing sweep. This demonstrates that the multidomain state is strongly hysteretic and metastable, with the real-space configuration depending sensitively on the sweep history.

The comparison with the inset THE loop shows that the largest anomalies in $\rho_{xy}^{\mathrm{T}}$ occur precisely in the field regions where the MFM images exhibit the strongest texture modulation on each branch. Importantly, however, the relevant field windows differ between the two sweep directions. This indicates that the branch asymmetry of the THE is not a simple intrinsic difference between positive and negative magnetic fields, but is instead closely tied to the hysteretic formation, annihilation, and rearrangement of non-coplanar textures across the metamagnetic transition. Supplementary Figure S10 therefore provides direct real-space support for the conclusion that the magnitude and sign of $\rho_{xy}^{\mathrm{T}}$ are governed by the specific domain configuration stabilized along each field path.

**Expected MFM contrast and interpretation of the observed domain textures.**

To help interpret the MFM data presented in the main text, we summarize in Figure S11 the expected MFM contrast for several representative magnetic textures and discuss its implications for the present system. Because magnetic force microscopy probes the out-of-plane stray field above the sample surface, it is primarily sensitive to the uncompensated net magnetic moment, rather than to the microscopic spin configuration of individual atoms or magnetic sublattices.

For a conventional ferromagnetic stripe or bubble domain pattern, the out-of-plane magnetization produces a clear contrast between regions of opposite polarity, giving rise to strong bright and dark MFM signals. In a partially compensated ferrimagnetic state hosting a skyrmion-like or chiral bubble texture, the net out-of-plane component remains finite, so the MFM signal is likewise expected to show a localized bubble-like contrast. By contrast, in an ideal fully compensated antiferromagnetic skyrmion, the stray fields generated by the two antiparallel sublattices nearly cancel, such that the expected MFM contrast becomes extremely weak or even absent. Thus, Figure S11 illustrates both the usefulness and the limitation of MFM: it can provide real-space evidence for non-coplanar textures carrying a finite uncompensated moment, but it cannot by itself uniquely determine whether the microscopic texture is ferromagnetic, ferrimagnetic, or perfectly antiferromagnetic.

Within this framework, the MFM contrast should be interpreted as reflecting the spatial distribution of the uncompensated out-of-plane component of the magnetization rather than directly imaging the full microscopic Néel texture. Therefore, while the data are consistent with skyrmion-like or chiral bubble textures in a partially compensated state, they do not constitute direct proof of perfectly compensated antiferromagnetic skyrmions. Establishing such a claim would require element-specific or sublattice-sensitive probes capable of directly resolving the magnetic order on each sublattice.

**Same-tip comparison of Co-doped $Fe_3GaTe_2$ bubble-like textures and Nb vortices.**

To better estimate the intrinsic size of the bubble-like textures observed in Co-doped $Fe_3GaTe_2$, we carried out a direct comparison with superconducting vortices in a Nb film measured using the same MFM tip in the same cooldown sequence. The Nb film was positioned immediately next to the Co-doped $Fe_3GaTe_2$ sample, and the Nb vortex image was acquired after the magnetic imaging of the Co-doped $Fe_3GaTe_2$ sample without changing the probe. Therefore, the two data sets were obtained with essentially the same tip geometry and magnetic state, allowing a direct empirical comparison of the apparent lateral broadening in the MFM images.

Figure S12 compares a representative bubble-like texture in Co-doped $Fe_3GaTe_2$ at $\mu_0 H = -0.75$ T with vortices measured in the Nb film. In the Nb control image, the apparent vortex diameter is approximately 2 μm, even though the intrinsic magnetic length scale of the vortex is much smaller, of order ~ 100 nm. This directly demonstrates the substantial broadening introduced by the long-range stray field and tip convolution under our MFM imaging conditions.

The bubble-like contrast in Co-doped $Fe_3GaTe_2$ shows an apparent lateral size comparable to that of the broadened Nb vortices, rather than several micrometers larger. Because the Co-doped $Fe_3GaTe_2$ image was acquired with the same probe and under similar lift-height conditions, this comparison indicates that the intrinsic size of the bubble-like texture is also likely much smaller than the apparent MFM diameter. Using the Nb vortices as an internal calibration, we infer that the intrinsic diameter of the bubble-like textures in Co-doped $Fe_3GaTe_2$ is most likely on the order of ~100 nm, i.e. a few hundred nanometers or less, rather than on the micrometer scale directly suggested by the raw MFM contrast.

This estimate should be regarded as semi-quantitative, since the stray-field profiles of superconducting vortices and magnetic bubbles are not identical. Nevertheless, the same-tip, same-cooldown comparison provides a realistic empirical bound on the extent of MFM broadening in the present experiment. Importantly, it shows that the observed apparent size of the bubble-like features is compatible with nanoscale bubble textures once the unavoidable stray-field broadening is taken into account.

ASSOCIATED CONTENT

**Data Availability Statement**

The data underlying this study are available in the published article, its Supporting Information, and in the Open-Source Framework (https://osf.io/), through the following digital object identifier https://doi.org/10.17605/OSF.IO/EFJPM.

**Supporting Information**.

The Supporting Information is available free of charge at https://pubs.acs.org/doi/.

Crystallographic Data, Data Collection, and Refinement Parameters of $Fe_{3-x}Co_xGaTe_2$ (T = 298 K), Fractional Atomic Coordinates for $Fe_{3-x}Co_xGaTe_2$, Electric transport and magnetic properties of $Fe_{2.8}Co_{0.12}GaTe_2$, Energy dispersive X-ray spectrum (EDS) for $Fe_{2.4}Co_{0.6}GaTe_2$, Single crystal X-ray diffraction pattern from $Fe_{2.48}Co_{0.56}GaTe_2$, Resistivity and magnetization of a $Fe_{2.48}Co_{0.56}GaTe_2$ single-crystal as a function of the temperature for magnetic fields along the *ab*-plane, Anomalous Hall conductivity $\sigma_{xy}^{A}$ for $Fe_{3-x}Co_xGaTe_2$ crystals having different Co concentrations ($x$ = 0.12, 0.38, 0.56, and 0.91) at different $T$s, Magnetic Force Microscopy (MFM) images of nominal $Fe_{2.9}Co_{0.1}GaTe_2$, topological Hall effect extracted from a second $Co_{0.56}Fe_{2.44}GaTe_2$ crystal, image-processing pipeline used to extract quantitative domain metrics from the MFM images, sweep-direction dependence of the magnetic domain texture and topological Hall response, schematic comparison of expected MFM contrasts for different magnetic textures,

**s**ame-tip comparison between bubble-like textures in Co-doped $Fe_3GaTe_2$ and superconducting vortices in a Nb film

**Author Contributions.**

‡These authors contributed equally to this work. This manuscript was written through contributions of all authors. All authors have given approval to the final version of the manuscript.

ACKNOWLEDGMENTS

L.B. acknowledges support from the US DoE, BES program, through award DE-SC0002613 and start-up funds from Baylor University. J.Y.C. acknowledges NSF DMR-2505304 and the Welch Foundation through AA-2056-20240404. J. K. acknowledges support from the National Research Foundation of Korea (NRF) grant funded by the Korean government (MSTI) (No. RS-2024-00410027). The National High Magnetic Field Laboratory acknowledges support from the US-NSF Cooperative agreement Grant DMR-2128556, and the state of Florida.

Supplementary Information for

# Topological Hall Effect in Antiferromagnetic Co doped $Fe_3GaTe_2$

*Shyam Raj Karullithodi[1, 2, ‡], Yeonkyu Lee[3, 4, ‡], Vadym Kulichenko[1], W. Kice Brown[5], Sang-Eon Lee[1], Chanyoung Lee[3, 4], Jinyoung Yun[3, 4], Gregory T. McCandless[5], Julia Y. Chan[5], Jeehoon Kim[3, 4, *], and Luis Balicas[6, *]*

1. National High Magnetic Field Laboratory, Tallahassee, Florida 32310, United States
2. Department of Physics, Florida State University, Tallahassee, Florida 32306, United States
3. Department of Physics, Pohang University of Science and Technology, Pohang 37673, Republic of Korea
4. Center for Quantum Dynamics of Angular Momentum, Pohang University of Science and Technology, Pohang 37673, Republic of Korea
5. Department of Chemistry & Biochemistry, Baylor University, Waco, Texas 76706, United States
6. Department of Physics and Astronomy, Baylor University, Waco, Texas 76706, United States

*Correspondence to E-mail: jeehoon@postech.ac.kr; luis_balicas@baylor.edu

**Methods.**

**Figure S1**. Comparison among the lattice parameters of the $Fe_{3-x}Co_xGaTe_2$ series where the nominal concentration (top) is compared to the one obtained from elemental analysis (bottom).

**Table S1**. Crystallographic Data, Data Collection, and Refinement Parameters of $Fe_{3-x}Co_xGaTe_2$ ($T$ = 298 K).

**Table S2**. Fractional Atomic Coordinates for $Fe_{3-x}Co_xGaTe_2$.

**Figure S2**. Electric transport and magnetic properties of $Fe_{2.8}Co_{0.2}$GaTe2.

**Figure S3**. Energy dispersive X-ray spectrum (EDS) for $Fe_{2.48}Co_{0.56}GaTe_2$.

**Figure S4**. Single crystal X-ray diffraction pattern from $Fe_{2.48}Co_{0.56}GaTe_2$.

**Figure S5.** Resistivity and magnetization for a $Fe_{2.48}Co_{0.56}GaTe_2$ single-crystal as a function of the temperature for magnetic fields along the *ab*-plane.

**Figure S6**. Anomalous Hall conductivity $\sigma_{xy}^{A}$ for $Fe_{2-x}Co_xGaTe_2$ crystals having different Co concentrations ($x$ = 0.12, 0.38, 0.56, and 0.91) at different $T$s.

**Figure S7**. Magnetic force microscopy (MFM) images from $Fe_{2.9}Co_{0.1}GaTe_2$ (nominal Co concentration).

**Figure S8**. Topological Hall effect extracted from a second $Co_{0.56}Fe_{2.44}GaTe_2$ crystal

**Figure S9**. Image-processing procedure used to extract quantitative domain metrics from the MFM images.

**Figure S10**. Sweep-direction dependence of the magnetic domain texture and topological Hall response.

**Figure S11.** Schematic comparison of expected MFM contrasts for different magnetic textures.

**Figure S12**. Same-tip comparison between bubble-like textures in Co-doped $Fe_3GaTe_2$ and superconducting vortices in a Nb film

## METHODS

### Crystal growth.

Single crystals of Co doped $Fe_{3-x}Co_xGaTe_2$ were synthesized through a chemical vapor transport method. High purity Fe and Co powders, Ga lumps, and Te lumps were mixed in the molar ratio (3-$x$):$x$:1:2 and sealed in an evacuated quartz tube. The mixture was pre-reacted at 1000 ˚C for 24 hours. The resulting product was finely ground in an agate mortar. The pre-reacted powder was sealed in a vacuum quartz tube with a small amount of iodine (5 mg/cm$^3$). The tube was then placed in a two-zone horizontal tube furnace with a temperature gradient of 750 - 720 ˚C for two weeks. The tube was quenched into ice-water. The resulting crystals were washed with acetone and isopropanol.

### Single crystal X-ray diffraction.

Single-crystal X diffraction data were obtained at room temperature using small crystals of $Fe_{2-x}Co_xGaTe_2$. Data was collected using a Bruker D8 Quest Kappa single-crystal X-ray diffractometer equipped with an Incoatec IμS microfocus source (Mo Kα radiation, λ = 0.71073 Å) and a PHOTON III CPAD area detector. The raw frames were integrated with Bruker SAINT, and the intensities were corrected for absorption with a multi-scan method in SADABS 2016/2.[32] The intrinsic phasing method in SHELXT was used to generate preliminary crystallographic models[33], which were finalized with least squares refinements in SHELXL2019.[34] The details for the crystallographic data, data collection, and the refinement parameters are outlined in Table S1 below. The crystallographic structure of $Fe_{3-x}Co_xGaTe_2$ (for $x$ = 0.12, 0.38, 0.56, 0.6, 0.91, and 1.05) consists of five atomic sites as outlined in Table S2. There is a contraction of the lattice parameters (~0.025 Å along $a$, ~0.05 Å along $c$, and ~3 Å$^3$ in volume) with increasing Co concentration (depicted in Figure S1).

**Electrical transport and magnetization measurements.**

The electrical transport measurements were performed in a Physical Property Measurement System (Quantum Design PPMS). The resistivity of the crystals was measured using a four-probe method. Magnetoresistance and Hall resistance were measured from $T$ = 300 K to 1.8 K by sweeping the magnetic field from $\mu_0 H$ = -9 T to $\mu_0 H$ = 9 T. DC magnetization $M$ was collected in a Magnetic Property Measurement System (Quantum Design MPMS-XL). Field cooled (FC) and zero field cooled (ZFC) measurements were measured from $T$ = 300 K to $T$ = 1.8 K. The magnetization of the sample was measured as a function of external magnetic field for values ranging from $\mu_0 H$ = -7 T to $\mu_0 H$ = 7 T.

**Magnetic force microscopy.**

The individual magnetic domains of the $Fe_{2.48}Co_{0.56}GaTe_2$ crystals were imaged using a home-built Magnetic Force Microscopy (MFM) at $T$ = 4.2 K. While Figure 4 describes the texture evolution during a single monotonic field sweep from +7 T to -7 T, Figure S10 presents the extended MFM datasets covering both sweep directions: from +7 T to -7 T and the reverse path from -7 T to +7 T. The system is completely polarized under 7 T. Images were taken in certain field intervals upon varying the field. To clarify how the real-space magnetic textures give rise to the topological Hall effect (THE) in this material, we quantitatively analyzed MFM images taken as a function of out-of-plane magnetic field $\mu_0 H$ at $T$ = 4.2 K and compared them with the topological Hall resistivity $\rho_{xy}^{\mathrm{T}}$ ($\mu_0 H$) measured at 1.8 K.

**Quantitative analysis of MFM domain morphology.**

To quantify the evolution of the magnetic domain structure as a function of magnetic field, we analyzed each MFM image using a common image-processing pipeline. Figure S9 illustrates the procedure for a representative field.

1) Normalization and smoothing. The raw MFM frequency-shift image is converted to a grayscale matrix and normalized to the range [0,1]. A mild Gaussian filter (typical kernel $\sigma$ = 1–2 pixels) is applied to suppress high-frequency noise without altering the overall domain morphology.

2) Binary segmentation and up-domain fraction $f_{\uparrow}$. An intensity threshold is determined by Otsu's method, which maximizes the inter-class variance between "bright" and "dark" pixels. Pixels above (below) the threshold are assigned to up (down) domains, yielding a binary map [Fig. 5(b)]. The up-domain area fraction is then computed as $f_{\uparrow} = N_{up} /(N_{up} + N_{down})$, where $N_{up}$ and $N_{down}$ are the number of pixels in each class.

3) Domain-wall extraction and length $L_{DW}$. We apply a Canny edge detector to the smoothed grayscale image to extract the boundaries between bright and dark regions [Fig. 5(c)]. Small speckles consisting of fewer than $N_{min}$ connected pixels (typically $N_{min}$ = 10) are removed to avoid counting noise as domain walls. The remaining connected edge pixels form extended domain walls. The total domain wall length is then proportional to the number of edge pixels: $L_{DW} = a \times N_{edge}$, where a denotes the pixel size in real space. In the main text we plot $N_{edge}$ (in pixels) as a convenient measure of $L_{DW}$

4) Intensity variance. Finally, we compute the variance of the normalized intensity over the entire image, $\mathrm{Var(I)} = \langle I^2 \rangle - \langle I \rangle^2$, which quantifies the overall domain contrast and mixture of bright and dark regions. Each of these metrics is evaluated for all MFM images and plotted as a function of magnetic field in Fig. 5(a)–(c) of the main text.

**Sweep-direction dependence of the MFM textures and topological Hall response.**

To further examine the branch asymmetry of the topological Hall effect, we performed additional MFM imaging for both decreasing- and increasing-field sweeps and compared the corresponding domain evolution with the field dependence of $\rho_{xy}^{\mathrm{T}}$ . The results are summarized in Supplementary Figure S10. In the corrected labeling of Fig. 3c, the decreasing-field sweep is denoted by $\mu_0 H_{dec}$ (blue), whereas the increasing-field sweep is denoted by $\mu_0 H_{inc}$ (magenta).

The MFM images clearly show that the magnetic textures do not retrace the same sequence upon reversing the field sweep direction. During the decreasing-field sweep from +7 T to −7 T, the system evolves from a nearly field-polarized state into a strongly modulated multidomain texture in two characteristic field windows highlighted by the red boxes: first around $\mu_0 H \approx +1.5$ to +1 T, and second around $\mu_0 H \approx -3$ to -3.25 T. The latter field range coincides closely with the strongest positive peak of $\rho_{xy}^{\mathrm{T}}$ observed on the $\mu_0 H_{dec}$ branch. These images show a dense stripe- and bubble-like morphology, suggesting that this branch stabilizes a particularly favorable non-coplanar texture for generating a large emergent field.

By contrast, during the increasing-field sweep from −7 T to +7 T, the magnetic texture evolves through a distinctly different sequence of states. The strongest multidomain contrast is observed in the red-boxed field windows around $\mu_0 H \approx -1.25$ to −0.75 T and $\mu_0 H \approx +3$ to +3.25 T, rather than in the same field ranges emphasized during the decreasing sweep. This demonstrates that the multidomain state is strongly hysteretic and metastable, with the real-space configuration depending sensitively on the sweep history.

The comparison with the inset THE loop shows that the largest anomalies in $\rho_{xy}^{\mathrm{T}}$ occur precisely in the field regions where the MFM images exhibit the strongest texture modulation on each branch. Importantly, however, the relevant field windows differ between the two sweep directions. This indicates that the branch asymmetry of the THE is not a simple intrinsic difference between positive and negative magnetic fields, but is instead closely tied to the hysteretic formation, annihilation, and rearrangement of non-coplanar textures across the metamagnetic transition. Supplementary Figure S10 therefore provides direct real-space support for the conclusion that the magnitude and sign of $\rho_{xy}^{\mathrm{T}}$ are governed by the specific domain configuration stabilized along each field path.

**Expected MFM contrast and interpretation of the observed domain textures.**

To help interpret the MFM data presented in the main text, we summarize in Figure S11 the expected MFM contrast for several representative magnetic textures and discuss its implications for the present system. Because magnetic force microscopy probes the out-of-plane stray field above the sample surface, it is primarily sensitive to the uncompensated net magnetic moment, rather than to the microscopic spin configuration of individual atoms or magnetic sublattices.

For a conventional ferromagnetic stripe or bubble domain pattern, the out-of-plane magnetization produces a clear contrast between regions of opposite polarity, giving rise to strong bright and dark

MFM signals. In a partially compensated ferrimagnetic state hosting a skyrmion-like or chiral bubble texture, the net out-of-plane component remains finite, so the MFM signal is likewise expected to show a localized bubble-like contrast. By contrast, in an ideal fully compensated antiferromagnetic skyrmion, the stray fields generated by the two antiparallel sublattices nearly cancel, such that the expected MFM contrast becomes extremely weak or even absent. Thus, Figure S11 illustrates both the usefulness and the limitation of MFM: it can provide real-space evidence for non-coplanar textures carrying a finite uncompensated moment, but it cannot by itself uniquely determine whether the microscopic texture is ferromagnetic, ferrimagnetic, or perfectly antiferromagnetic.

Within this framework, the MFM contrast should be interpreted as reflecting the spatial distribution of the uncompensated out-of-plane component of the magnetization rather than directly imaging the full microscopic Néel texture. Therefore, while the data are consistent with skyrmion-like or chiral bubble textures in a partially compensated state, they do not constitute direct proof of perfectly compensated antiferromagnetic skyrmions. Establishing such a claim would require element-specific or sublattice-sensitive probes capable of directly resolving the magnetic order on each sublattice.

**Same-tip comparison of Co-doped $Fe_3GaTe_2$ bubble-like textures and Nb vortices.**

To better estimate the intrinsic size of the bubble-like textures observed in Co-doped $Fe_3GaTe_2$, we carried out a direct comparison with superconducting vortices in a Nb film measured using the same MFM tip in the same cooldown sequence. The Nb film was positioned immediately next to the Co-doped $Fe_3GaTe_2$ sample, and the Nb vortex image was acquired after the magnetic imaging of the Co-doped $Fe_3GaTe_2$ sample without changing the probe. Therefore, the two data sets were obtained with essentially the same tip geometry and magnetic state, allowing a direct empirical comparison of the apparent lateral broadening in the MFM images.

Figure S12 compares a representative bubble-like texture in Co-doped $Fe_3GaTe_2$ at $\mu_0 H = -0.75$ T with vortices measured in the Nb film. In the Nb control image, the apparent vortex diameter is approximately 2 μm, even though the intrinsic magnetic length scale of the vortex is much smaller, of order ~ 100 nm. This directly demonstrates the substantial broadening introduced by the long-range stray field and tip convolution under our MFM imaging conditions.

The bubble-like contrast in Co-doped $Fe_3GaTe_2$ shows an apparent lateral size comparable to that of the broadened Nb vortices, rather than several micrometers larger. Because the Co-doped $Fe_3GaTe_2$ image was acquired with the same probe and under similar lift-height conditions, this comparison indicates that the intrinsic size of the bubble-like texture is also likely much smaller than the apparent MFM diameter. Using the Nb vortices as an internal calibration, we infer that the intrinsic diameter of the bubble-like textures in Co-doped $Fe_3GaTe_2$ is most likely on the order of ~100 nm, i.e. a few hundred nanometers or less, rather than on the micrometer scale directly suggested by the raw MFM contrast.

This estimate should be regarded as semi-quantitative, since the stray-field profiles of superconducting vortices and magnetic bubbles are not identical. Nevertheless, the same-tip, same-cooldown comparison provides a realistic empirical bound on the extent of MFM broadening in the present experiment. Importantly, it shows that the observed apparent size of the bubble-like features is compatible with nanoscale bubble textures once the unavoidable stray-field broadening is taken into account.

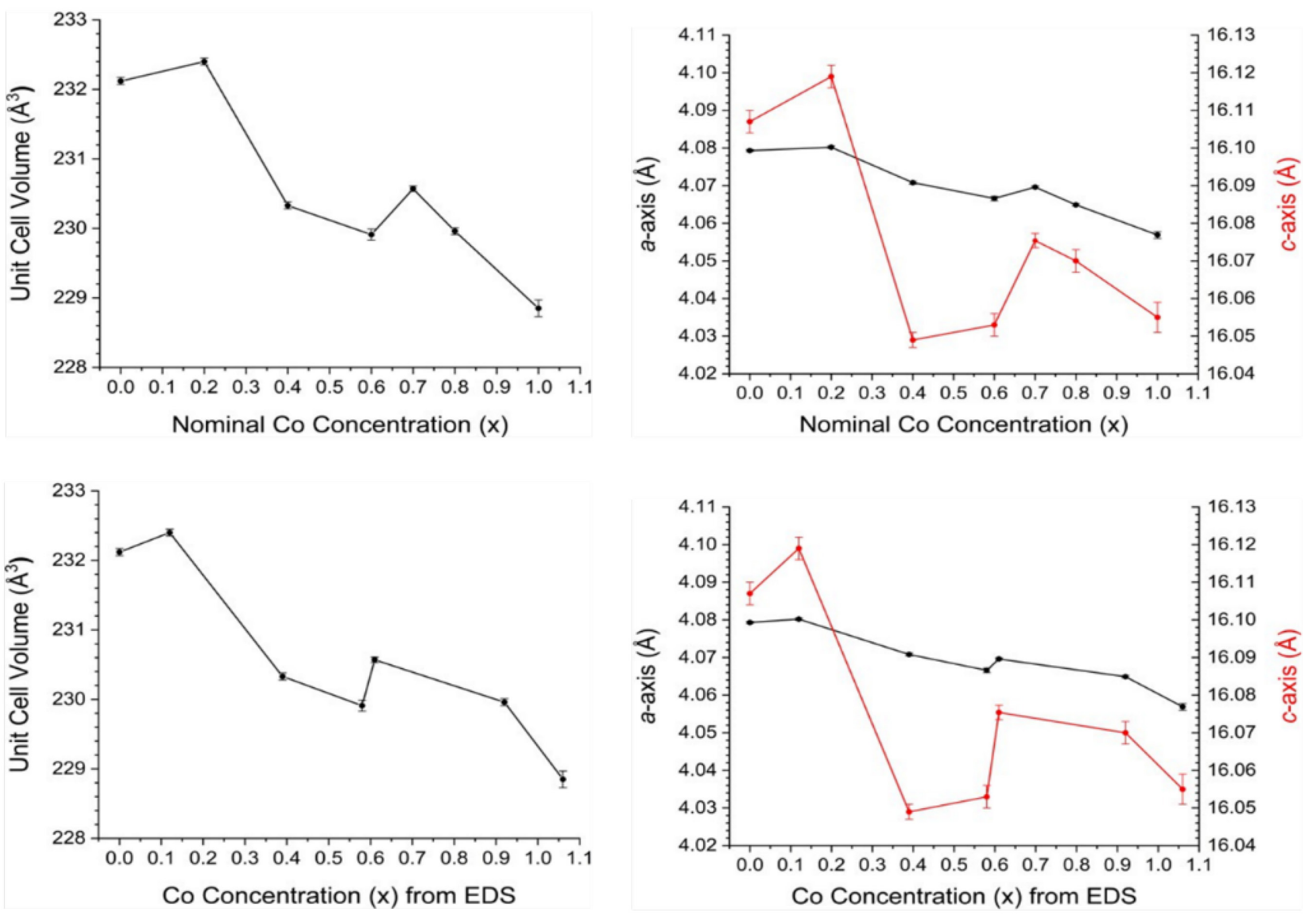
Unit Cell Volume (Å³)
Nominal Co Concentration (x)
a-axis (Å)
c-axis (Å)
Nominal Co Concentration (x)
Unit Cell Volume (Å³)
Co Concentration (x) from EDS
a-axis (Å)
c-axis (Å)
Co Concentration (x) from EDS

**Figure S1**. Comparison among the lattice parameters of the $Fe_{3-x}Co_xGaTe_2$ series where the nominal concentration (top) is compared to the one obtained from elemental analysis (bottom).

**Table S1. Crystallographic Data, Data Collection, and Refinement Parameters of $Fe_{3-x}Co_xGaTe_2$ ($T$ = 298 K)**

| Nominal Formula | $x$ = 0.2 | $x$ = 0.4 | $x$ = 0.6 | $x$ = 0.7 | $x$ = 0.8 | $x$ = 1.0 |
|---|---|---|---|---|---|---|
| Refined Formula* | $Fe_{2.89}Co_{0.12}GaTe_2$ | $Fe_{2.67}Co_{0.38}GaTe_2$ | $Fe_{2.48}Co_{0.56}GaTe_2$ | $Fe_{2.41}Co_{0.60}GaTe_2$ | $Fe_{2.13}Co_{0.91}GaTe_2$ | $Fe_{1.98}Co_{1.05}GaTe_2$ |
| Space Group | $P6_3/mmc$ | | | | | |
| $a$ (Å) | 4.0802(3) | 4.0708(4) | 4.0666(6) | 4.0691(7) | 4.0649(3) | 4.0569(9) |
| $c$ (Å) | 16.119(3) | 16.049(2) | 16.053(3) | 16.055(4) | 16.070(3) | 16.055(4) |
| $V$ (Å$^3$) | 232.40(5) | 230.33(5) | 229.91(8) | 230.22(10) | 229.96(5) | 228.85(12) |
| $Z$ | 2 | | | | | |
| $\Theta$ (°) | 2.5 – 30.5 | 2.5 – 30.6 | 2.5 – 30.5 | 2.5 – 30.6 | 2.5 – 30.5 | 2.5 – 30.5 |
| $\mu$ (mm$^{-1}$) | 27.10 | 27.52 | 27.69 | 27.57 | 27.83 | 28.00 |
| Measured Reflections | 8599 | 8170 | 8238 | 9477 | 8013 | 8399 |
| Independent Reflections | 173 | 172 | 170 | 173 | 171 | 170 |
| $R_{int}$ | 0.060 | 0.068 | 0.073 | 0.072 | 0.052 | 0.072 |
| $\Delta\rho_{max}$ (e-/Å$^3$) | 2.44 | 1.61 | 1.53 | 1.92 | 1.07 | 1.72 |
| $\Delta\rho_{min}$ (e-/Å$^3$) | -2.08 | -1.78 | -1.91 | -2.18 | -1.19 | -1.59 |
| Extinction Coefficient | 0.037(4) | none | none | 0.019(5) | 0.0071(7) | 0.006(2) |
| $R_1$ [$F^2 > 2\sigma(F^2)$] | 0.034 | 0.030 | 0.028 | 0.060 | 0.017 | 0.030 |
|  | 0.078 | 0.072 | 0.070 | 0.156 | 0.040 | 0.074 |

$R_1 = \sum||F_o| - |F_c|| / \sum|F_o|$
$wR_2 = (\sum w[(F_o)^2 - (F_c)^2]^2 / \sum w [(F_o)^2]^2)^{1/2}$

*Refined formulae are based on constrained Fe:Co ratios obtained from EDS values.

**Table S2. Fractional Atomic Coordinates for $Fe_{3-x}Co_xGaTe_2$**

| Site | Site Symmetry | *x* | *y* | *z* | Occ. |
|---|---|---|---|---|---|
| Fe/Co1 | 3 *m* | 0 | 0 | ~0.67 | 1 |
| Fe/Co2 | -6 *m* 2 | 2/3 | 1/3 | 0.75 | ~0.9 |
| Fe/Co3 | 3 *m* | 0 | 0 | 0.5 | ~0.1 |
| Ga1 | -6 *m* 2 | 1/3 | 2/3 | 0.75 | 1 |
| Te1 | 3 *m* | 2/3 | 1/3 | ~0.59 | 1 |

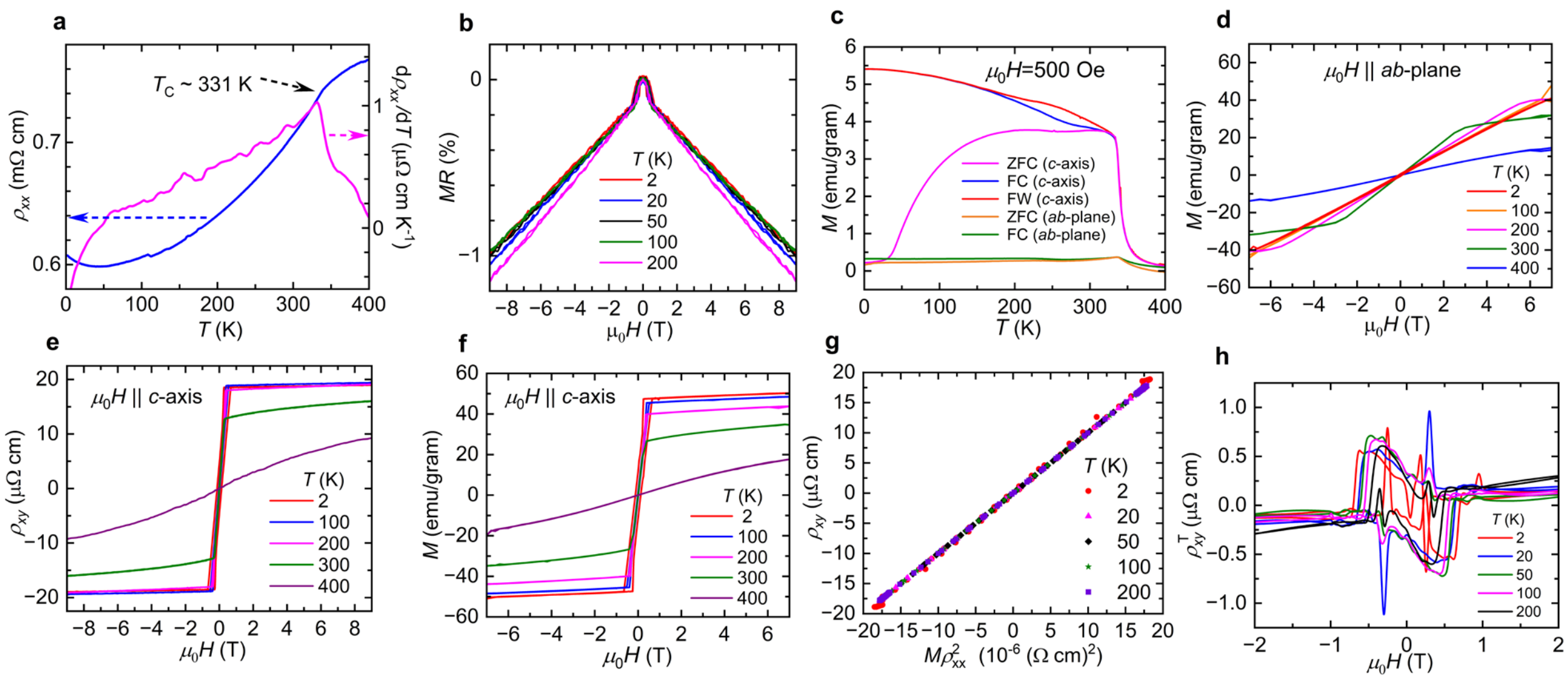

**Figure S2**. Transport and magnetic properties of $Fe_{2.48}Co_{0.56}GaTe_2$. (a) Resistivity $\rho_{xx}$ as a function of the temperature $T$ for a $Fe_{2.48}Co_{0.56}GaTe_2$ crystal (left axis) and its first derivative with respect to $T$ to expose the magnetic transition (right axis). (b) Magneto resistivity at different temperatures. (c) Magnetization as a function of temperature, along $c$-axis and $ab$-plane. (d) Magnetization $M$ as a function of external magnetic field $\mu_0H$ along $ab$-plane of the crystal. (e) Anomalous Hall effect $\rho_{xy}$ measured at different $T$s for fields applied along the $c$-axis of the crystal. (f) $M$ as a function of $\mu_0H$ applied along the $c$-axis. (g) $\rho_{xy}$ as a function of the product of $M$ and $\rho_{xy}^2$. (h) Topological Hall signal, once the anomalous Hall is subtracted, at several temperatures.

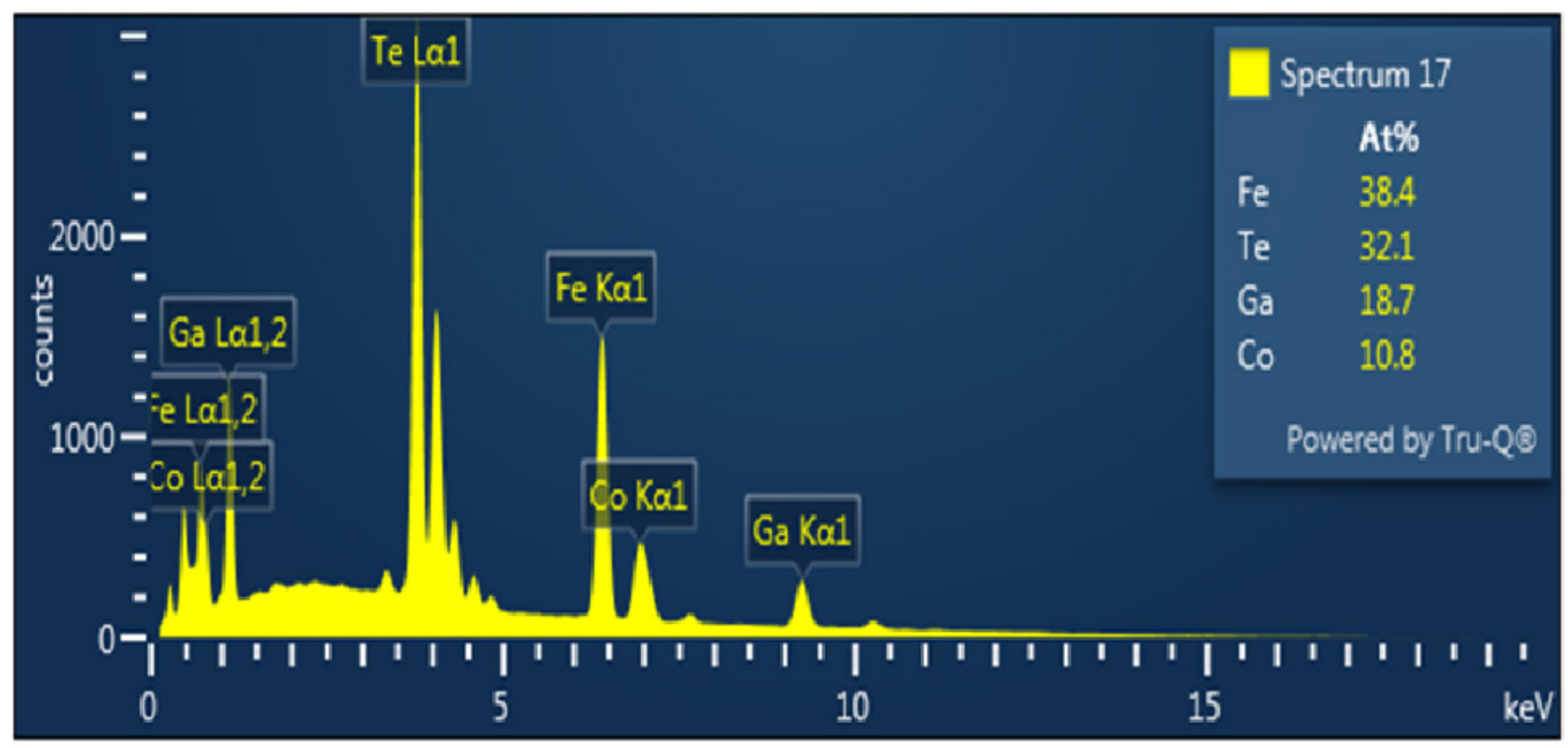

**Figure S3**. Energy dispersive X-ray spectrum (EDS) for nominal $Fe_{2.48}Co_{0.56}GaTe_2$. EDS spectrum showing the atomic percentages for each compositional element.

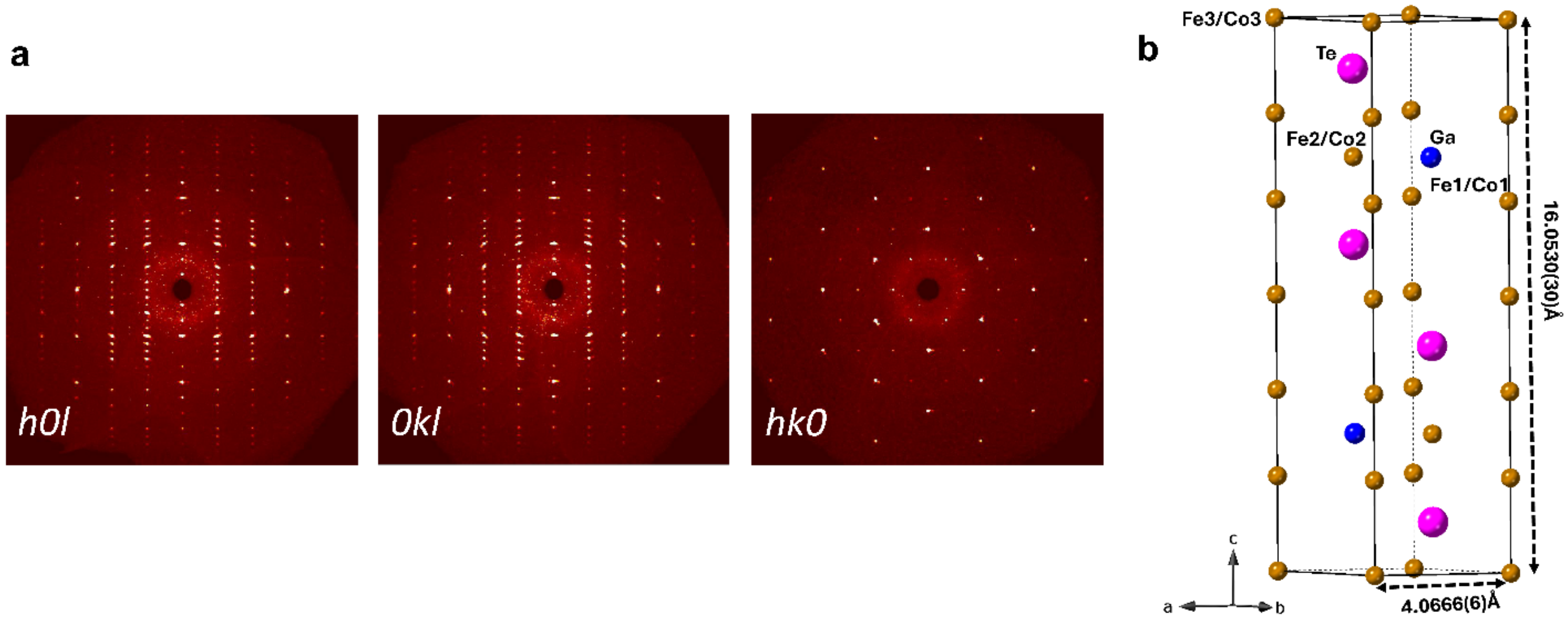

**Figure S4**. Single crystal X-ray diffraction pattern from $Fe_{2.48}Co_{0.56}GaTe_2$. (a) Indexed precession images from single crystal X-ray diffraction collected along *h0l*, *0kl*, and *hk0* planes, respectively, which are generated from the same data used to determine the crystal structure of $Fe_{2.48}Co_{0.56}GaTe_2$. (b) Unit cell representation of the crystal structure of $Fe_{2.48}Co_{0.56}GaTe_2$.

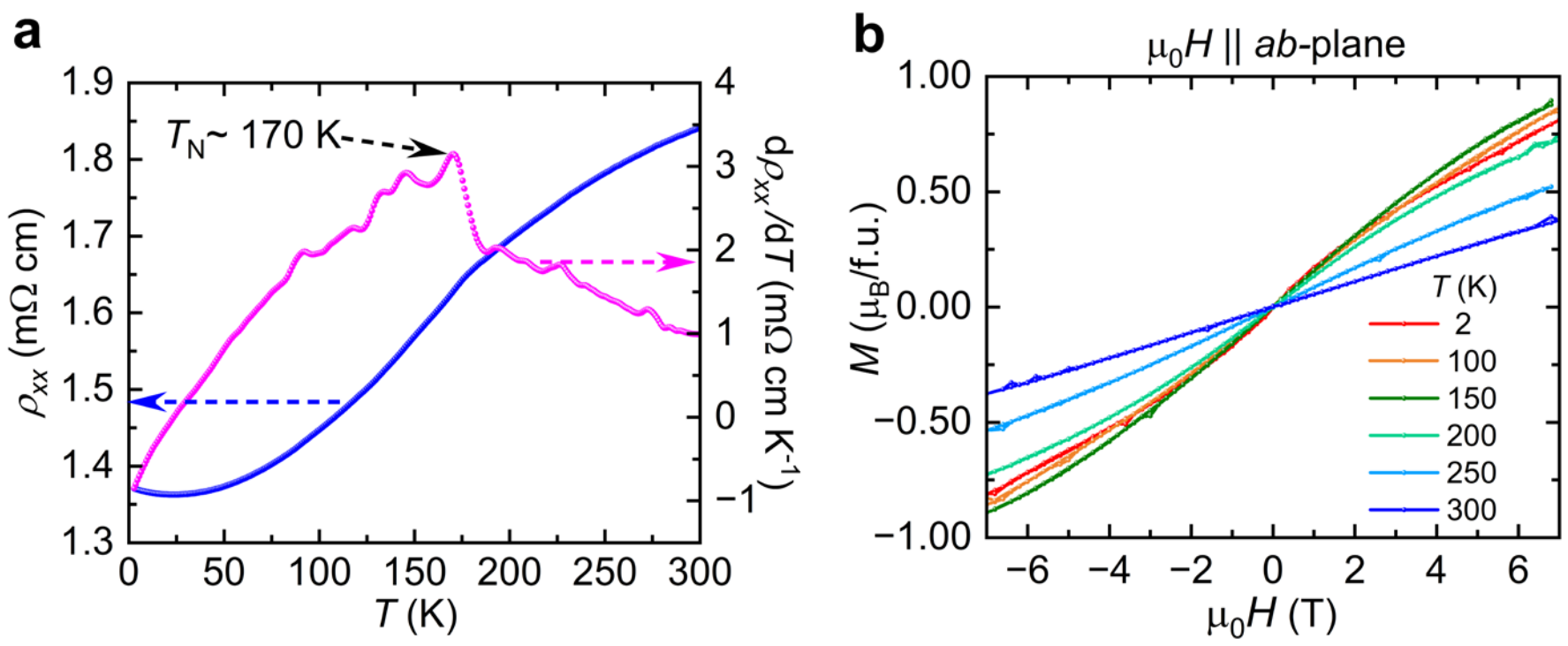

**Figure S5.** Resistivity and magnetization for a $Fe_{2.48}Co_{0.56}GaTe_2$ single-crystal as a function of the temperature for magnetic fields along the *ab*-plane. (a) In-plane resistivity $\rho_{xx}$ for a $Fe_{2.4}Co_{0.6}GaTe_2$ single-crystal as a function of $T$ (blue trace) with its derivative (magenta trace) clearly exposing a magnetic transition at $T_N \sim 170$ K. (b) $M$ as a function of $\mu_0H$ for different temperatures when the external magnetic field is applied along the *ab*-plane of the sample.

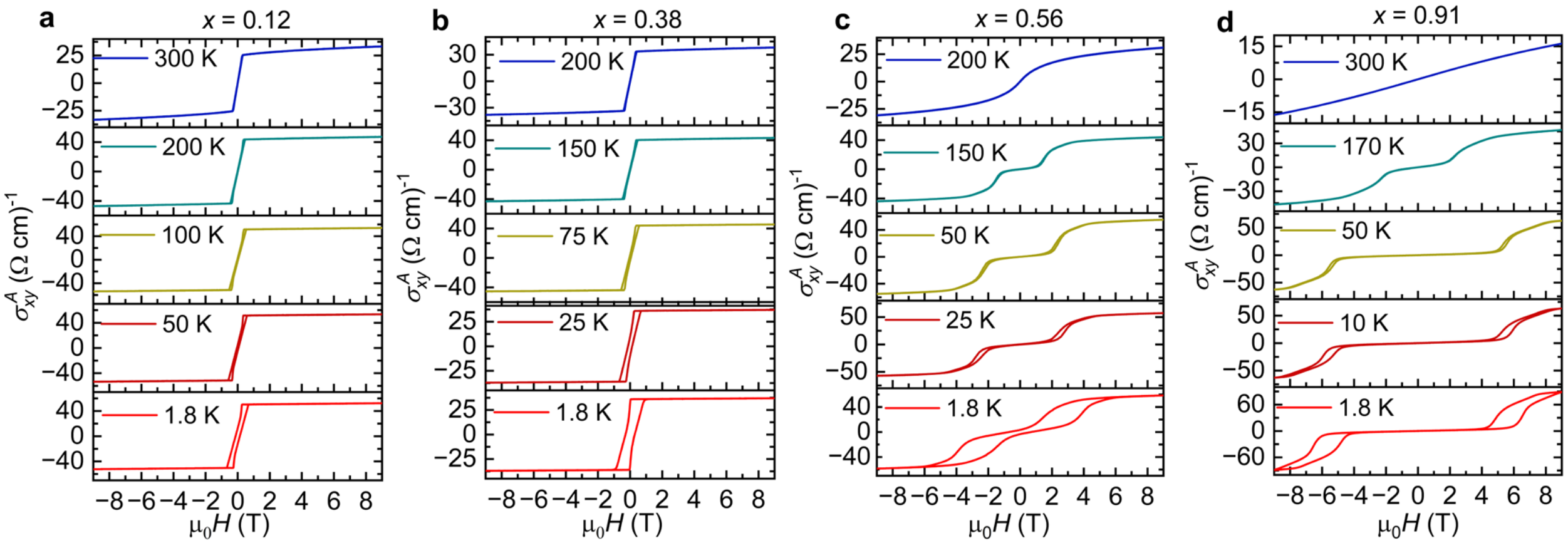

**Figure S6**. Anomalous Hall conductivity $\sigma_{xy}^A$ for $Fe_{3-x}Co_xGaTe_2$ crystals having different Co concentrations ($x$ = 0.12, 0.38, 0.56, and 0.91) at different $T$s. (a) For $x$ = 0.12, the magnetic coercive field is larger relative to the values observed in the parent $Fe_3GaTe_2$ compound. (b) For $x$ = 0.38, one observes even higher coercive fields while the system still remains in the ferromagnetic state. (c) For $x$ = 0.56, $M$ exposes its antiferromagnetic nature, namely almost zero magnetization at low fields until a metamagnetic transition

is reached resulting from a spin re-orientation transition. (d) For $x = 0.91$, higher magnetic fields are required to induce the spin-reorientation transition.

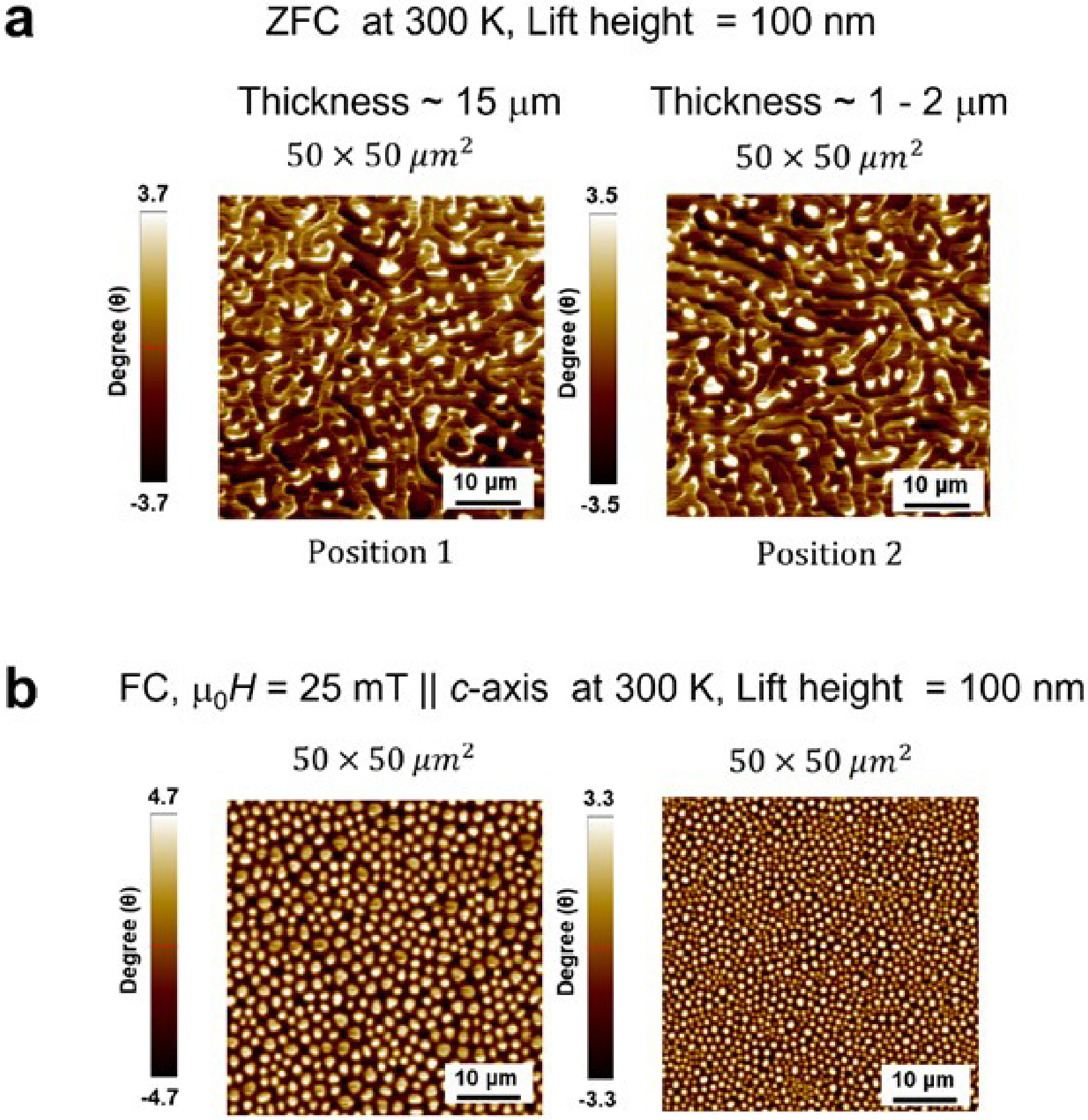

**Figure S7**. Magnetic force microscopy (MFM) images from $Fe_{2.9}Co_{0.1}GaTe_2$ (nominal Co concentration). (a) ZFC protocol reveals labyrinthine domains in MFM. (b) FC protocol leads to magnetic bubbles which are very likely to correspond to skyrmions.

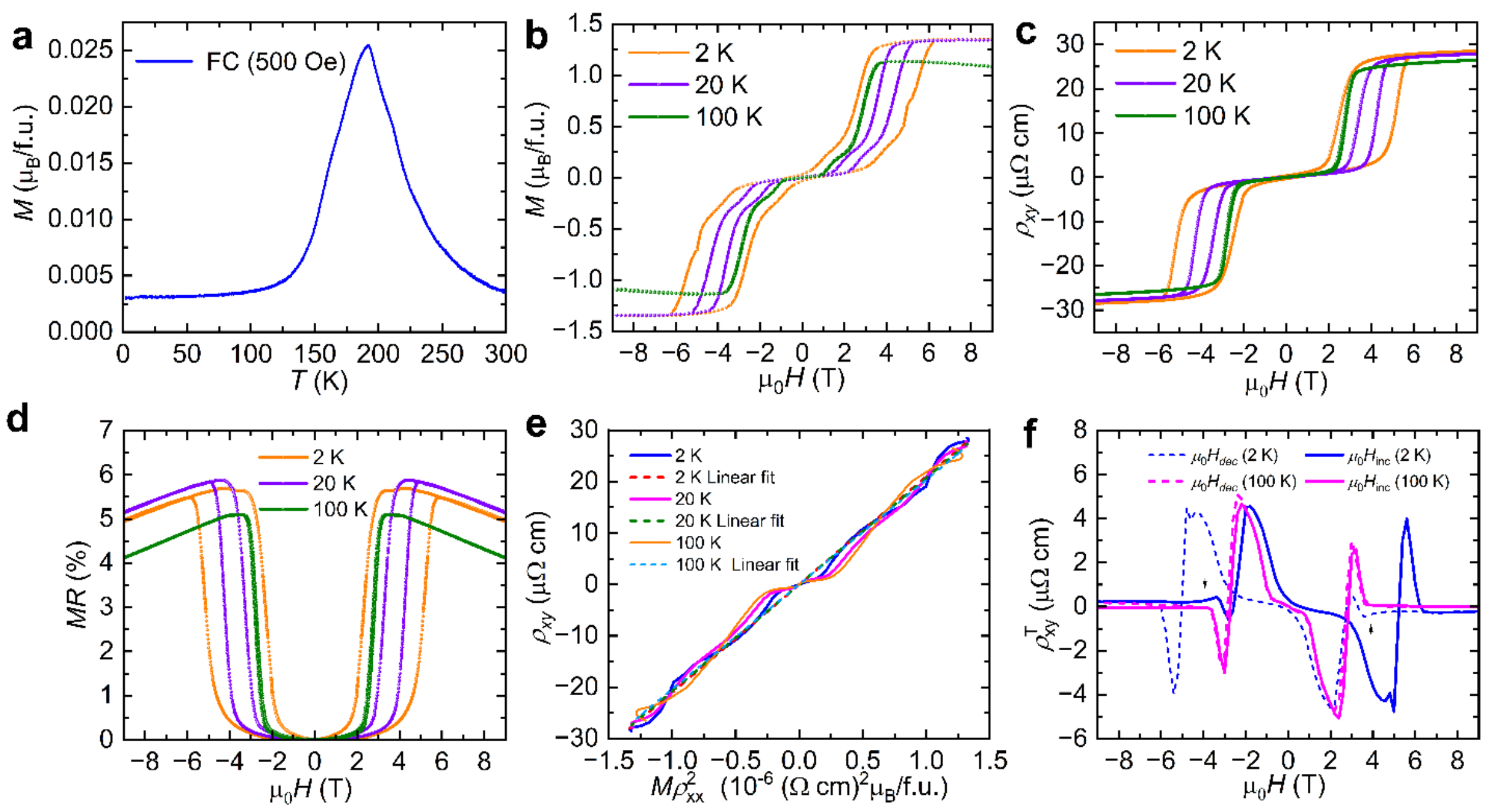

**Figure S8**. Topological Hall effect extracted from a second $Fe_{2.48}Co_{0.56}GaTe_2$ crystal. (a) Magnetization *M* as a function of the temperature for a second $Fe_{2.48}Co_{0.56}GaTe_2$ crystal for fields along *c*-axis. (b) Magnetization *M* as a function of external magnetic field $\mu_0 H$ applied along the *c*-axis of the crystal measured at $T = 2$ K, 20 K, and 100 K. (c) Anomalous Hall effect $\rho_{xy}$ measured on the same crystal at $T =$ 2 K, 20 K, and 100 K for fields along the *c*-axis of the crystal. (d) Magneto-resistivity at $T = 2$ K, 20 K, and 100 K. (e) $\rho_{xy}$ as a function of the product of *M* and $\rho_{xy}^2$. Dashed lines are linear fits. (f) Resulting topological Hall effect $\rho_{xy}^{\mathrm{T}}$, once the anomalous Hall signal is subtracted from the raw data, for $T = 2$ K (blue trace) and 100 K (magenta trace). Solid and dashed lines depict field increasing and decreasing sweeps, respectively. Notice that in this crystal the topological Hall signal is still observed at temperatures as high as $T = 100$ K.

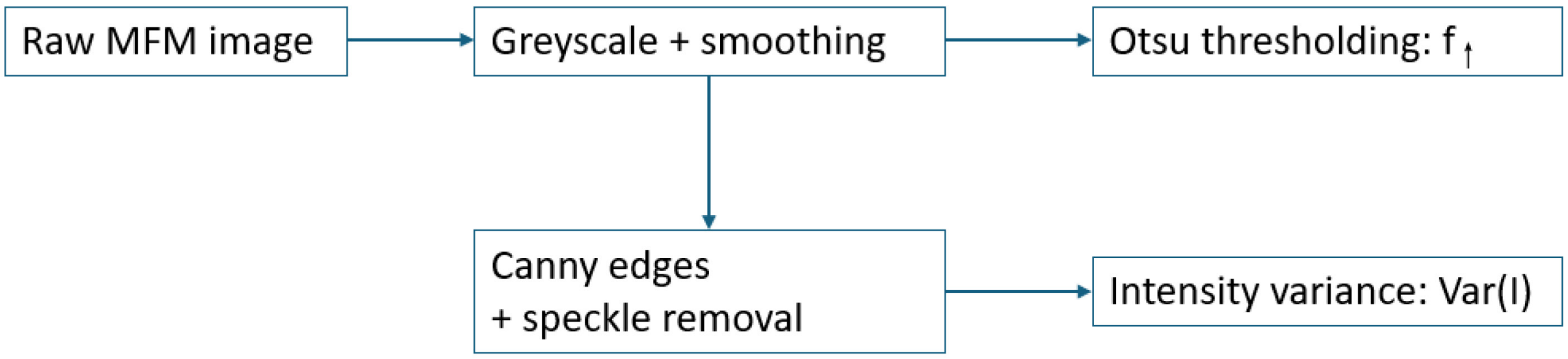

**Figure S9**. Image-processing procedure used to extract quantitative domain metrics from the MFM images. Each image is first normalized and smoothed, then segmented into up and down domains by Otsu thresholding to obtain the up-domain area fraction $f_{\uparrow}$. Domain walls are extracted using a Canny edge detector after removal of small speckles, and the total number of edge pixels provides a measure of the domain-wall length $L_{DW}$. The intensity variance Var(I) is computed directly from the normalized grayscale image. From these images we compute the metrics $f_{\uparrow}$, $L_{DW}$, and Var(I) used in Fig. 5 of the main text.

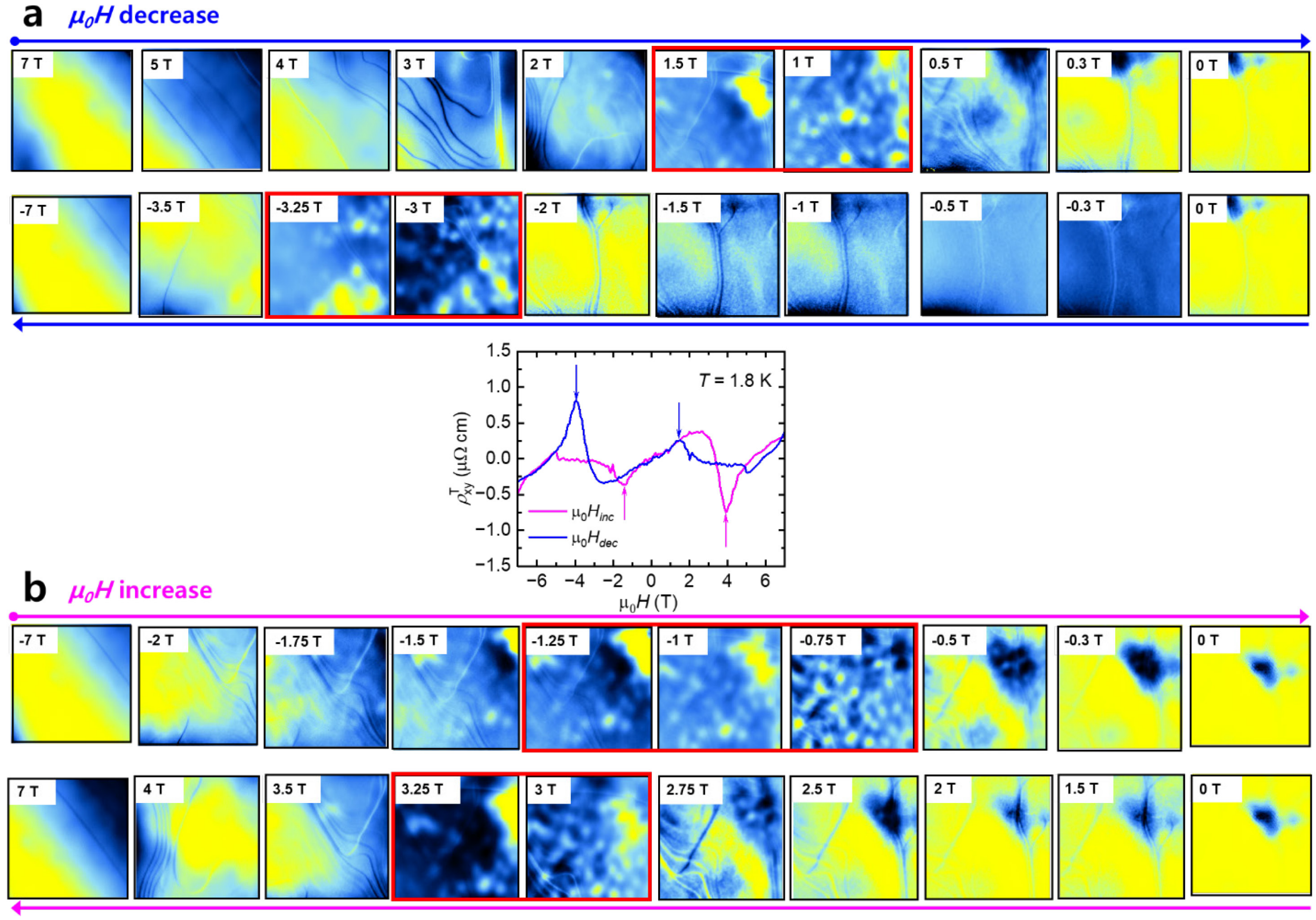

**Figure S10**. Dependence of the magnetic domain texture on sweep-direction and its influence on the topological Hall response. (a) MFM images acquired at $T = 4.2$ K during the decreasing-field sweep from $\mu_0 H = +7$ T to $-7$ T. (b) MFM images acquired during the increasing-field sweep from $\mu_0 H = -7$ T to +7 T. The colored arrows indicate the field sweep direction. Red boxes highlight field windows in which strongly modulated stripe- and bubble-like multidomain textures develop. The central inset shows the topological Hall resistivity $\rho_{xy}^{\mathrm{T}}$ at $T = 1.8$ K, with the corrected labels $\mu_0 H_{dec}$ (blue) and $\mu_0 H_{inc}$ (magenta).

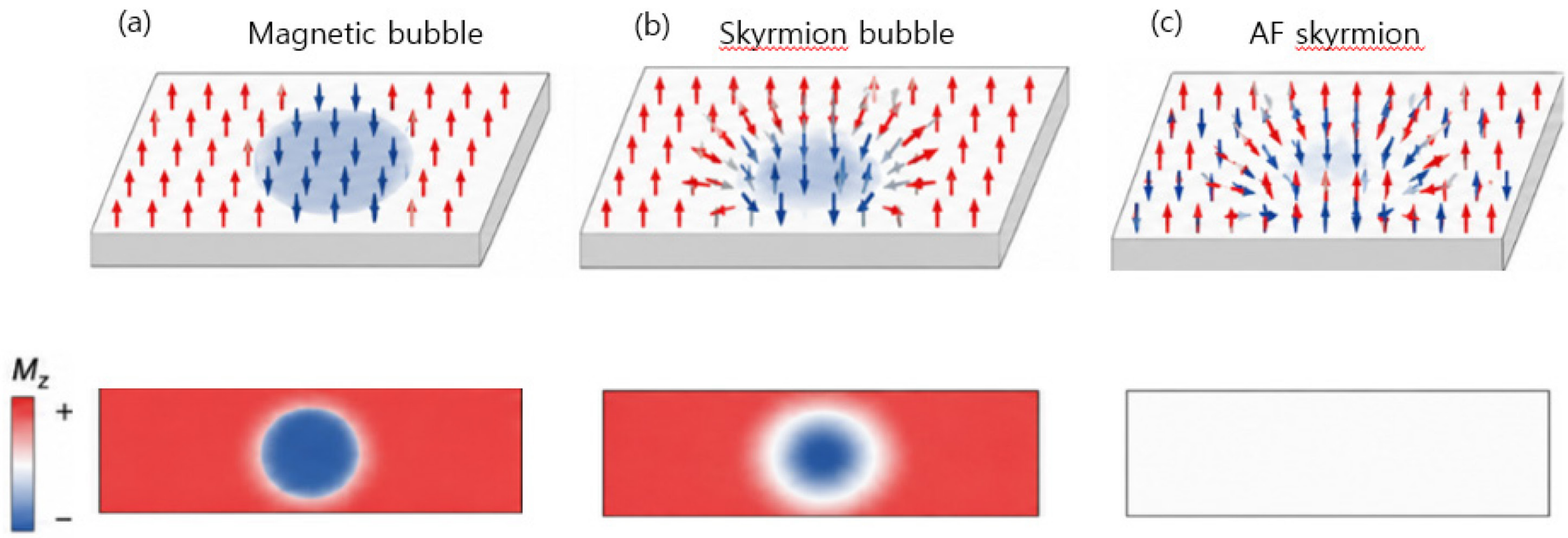

**Figure S11**. Schematic comparison of expected MFM contrasts for different magnetic textures **(a)** Conventional magnetic bubble in a ferromagnetic background. **(b)** Skyrmion-like bubble with a chiral spin winding in a partially compensated ferrimagnetic state. **(c)** Ideal antiferromagnetic skyrmion with nearly compensated sublattice moments. The upper row shows schematic spin configurations, and the lower row shows the corresponding out-of-plane magnetization $M_z$. While both the ferromagnetic bubble and the skyrmion-like bubble produce a finite $M_z$ contrast, an ideal compensated antiferromagnetic skyrmion yields nearly zero net $M_z$ and therefore would be expected to generate little or no detectable MFM signal. Therefore, the bubbles observed via MFM could result from partially spin polarized antiferromagnetic skyrmions.

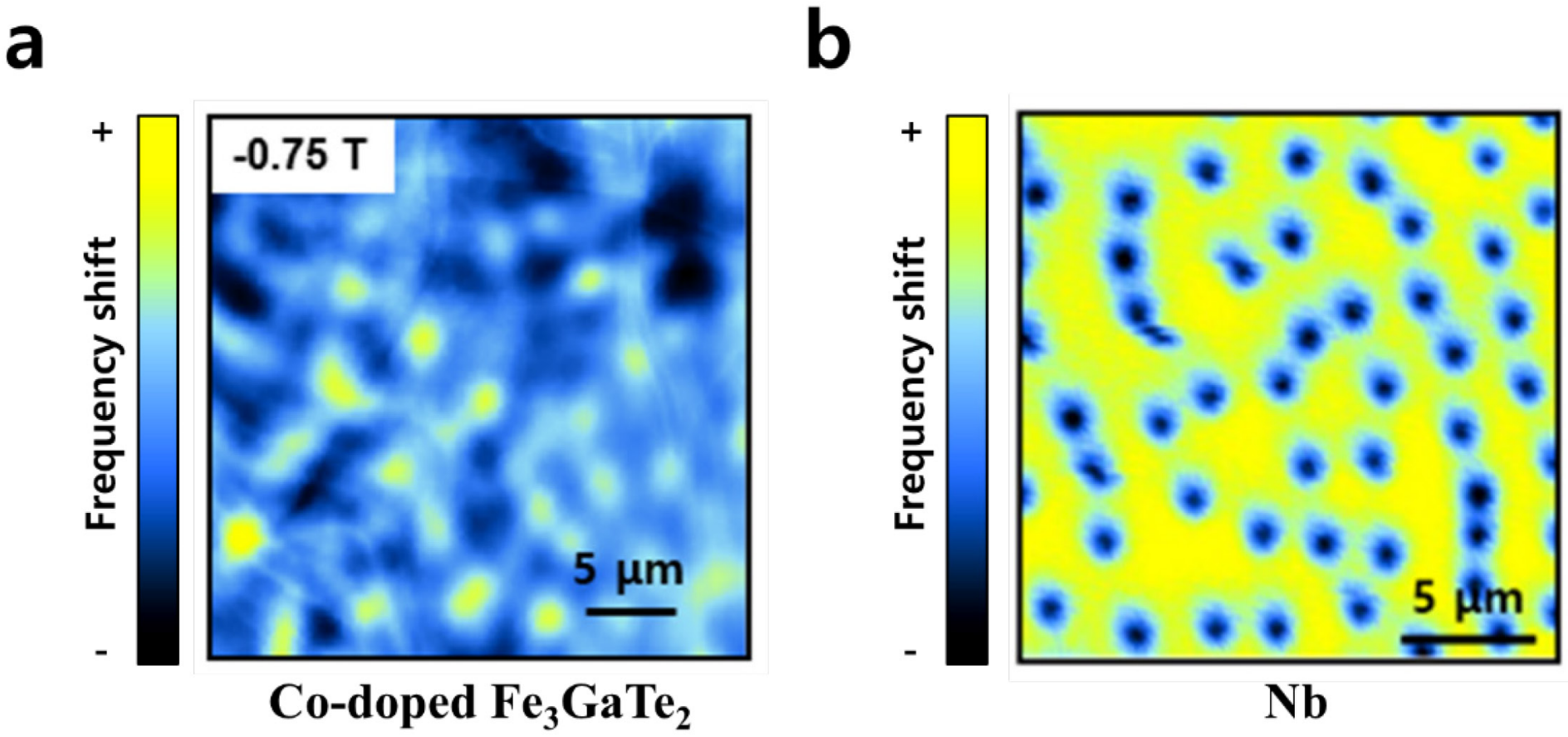

**Figure S12**. Same-tip comparison between bubble-like textures in Co-doped $Fe_3GaTe_2$ and superconducting vortices in a Nb film, measured at $T = 4.2$ K. (a) Representative MFM image of a bubble-like texture in Co-doped $Fe_3GaTe_2$ measured at $\mu_0 H = -0.75$ T. (b) MFM image of vortices in a Nb film acquired using the same tip in the same cooldown sequence after imaging the Co-doped $Fe_3GaTe_2$ sample.

**References-**